\documentclass[journal]{IEEEtran}

\usepackage[nocompress]{cite}

\ifCLASSINFOpdf
   \usepackage[pdftex]{graphicx}
  \else
  \usepackage[dvips]{graphicx}

\fi

\usepackage{amsmath}

\usepackage{array}
\usepackage{subfigure}
\usepackage{algorithm}
\usepackage{hyperref}
\usepackage{amssymb}
\usepackage{algorithmic}

\usepackage[1]{editing}

\begin{document}
%
\title{Covariance Matrix Construction with Preprocessing-Based Spatial Sampling for Robust Adaptive Beamforming }

\author{Saeed~Mohammadzadeh,~\IEEEmembership{Member,~IEEE,} Rodrigo~C.~de~Lamare,~\IEEEmembership{Senior Member,~IEEE,}
         and~Yuriy~Zakharov,~\IEEEmembership{Senior Member,~IEEE,}
        \vspace{-0.75em}}

\maketitle

\begin{abstract}
This work proposes an efficient, robust adaptive beamforming technique to deal with steering vector (SV) estimation mismatches and data covariance matrix reconstruction problems. In particular, the direction-of-arrival(DoA) of interfering sources is estimated with available snapshots in which the angular sectors of the interfering signals are computed adaptively. Then, we utilize the well-known general linear combination algorithm to reconstruct the interference-plus-noise covariance (IPNC) matrix using preprocessing-based spatial sampling (PPBSS). We demonstrate that the preprocessing matrix can be replaced by the sample covariance matrix (SCM) in the shrinkage method. A power spectrum sampling strategy is then devised based on a preprocessing matrix computed with the estimated angular sectors' information. Moreover, the covariance matrix for the signal is formed for the angular sector of the signal-of-interest (SOI), which allows for calculating an SV for the SOI using the power method. An analysis of the array beampattern in the proposed PPBSS technique is carried out, and a study of the computational cost of competing approaches is conducted. Simulation results show the proposed method's effectiveness compared to existing approaches.
\end{abstract}

%
\IEEEpeerreviewmaketitle

\section{Introduction}
\IEEEPARstart{A}{daptive} beamforming spans across various fields, including wireless communications, radar, sonar, and medical imaging, where it significantly improves performance by increasing signal-to-noise ratio (SNR) and mitigating interference \cite{van2004detection}. However, the beamforming performance degrades substantially under non-ideal conditions, such as finite data samples and mismatches between the assumed and actual steering vectors(SVs).
Several robust adaptive beamforming techniques have been developed to address model mismatches and enhance the robustness of beamformers. These methods generally fall into four categories: diagonal loading (DL), eigenspace-based approaches, uncertainty-set-based techniques, and approaches based on reconstructing the interference-plus-noise covariance (IPNC) matrix.
\subsection{Prior and Related Works}
\textit{Diagonal Loading Techniques:} These techniques enhance robustness against desired signal mismatches and the effects of limited training samples by incorporating a loading factor into the diagonal elements of the sample covariance matrix. However, its main drawback is that choosing the optimal DL factor in different scenarios is challenging \cite{li2003robust,elnashar2006further, mestre2006finite,jio,du2009review,du2010fully,yang2010modified,nai2011iterative, gu2011robust,kukrer2014generalised,rdrcb,xu2015response,l1stap,rmbthp,rdrls,dnlmsmest,rrdstap,yang2023robust,rdmt,rapa,rrsprec,rracf}.\\ 
\indent \textit{Eigenspace-Based Techniques :} This robust adaptive beamforming approach is based on projecting the nominal SVs onto the signal-plus-interference subspace to eliminate the arbitrary SV mismatches of the signal-of-interest (SOI). However, the performance of the eigenspace-based beamformer degrades drastically under low signal-to-noise ratios (SNR) \cite{chang1992performance,lee1997eigenspace,feldman1994projection, ccmmwf,jiostap,jidf,sjidf,khabbazibasmenj2012robust,wljio,dfjio,rralr,huang2012modified,ruan2014robust,oskpme, xie2014fast,rrber,yuan2017robust,mskaesprit,mohammadzadeh2018modified, lsomp,lrcc,rmmsecfprec,chen2024improved}.\\
\indent \textit{Uncertainty Set Techniques:} These techniques, such as the worst-case performance optimization and the linear programming algorithms, obtain an optimal solution by establishing an ellipsoidal uncertainty constraint on the SOI steering vector. Their performance is highly dependent on the uncertainty parameter set, which poses significant challenges in selecting optimal parameters in practical scenarios. Furthermore, these algorithms fail to exclude the SOI component from the sample covariance matrix, leading to significant performance degradation at high SNR levels \cite{vorobyov2003robust, vorobyov2008relationship,yu2010robust,jiang2014robust,yang2017modified,huang2023robust}.\\
\indent\textit{Interference-plus-Noise Covariance Matrix Reconstruction Techniques:} To address this issue, many works have focused on the removal of the signal-of-interest (SOI) components by reconstruction of the IPNC matrix instead of using the sample covariance matrix \cite{mallipeddi2011robust,gu2012robust,chen2015robust, huang2015robust, zhang2016interference,chen2018adaptive,mohammadzadeh2018adaptive,chen2018robust,mohammadzadeh2019robustcssp,zheng2018covariance,zhu2019covariance, Saeed2020,zhang2021adaptive, mohammadzadeh2021robust, Li2022Jul, Mohammadzadeh2022Dec, rsrbd,rsthp,rscf,luo2023urglq,9913641,9991136,li2023robust,luo2023urglq,chen2024robust,mohammadzadeh2022robust,mohammadzadeh2023jammer}. 
In \cite{mallipeddi2011robust}, the standard Capon beamformer was initially used to estimate the interference SV and reconstruct the IPNC matrix. However, it was found that the power of the interference and desired signal SVs are not accurately estimated. To address this, the Capon power spectrum was employed in \cite{gu2012robust} to reconstruct the IPNC matrix by integrating over an angle sector excluding that of the SOI. Additionally, the SOI was estimated by solving a quadratically constrained quadratic programming (QCQP) problem despite its high computational complexity. While this approach showed promising results, it was noted to be sensitive to large direction-of-arrival (DoA) mismatches, arbitrary amplitude, and phase perturbation errors \cite{yuan2017robust, mohammadzadeh2018modified}.\\ 
\indent In \cite{chen2015robust}, a correlation coefficient algorithm is used to construct the matrix, while the authors of \cite{huang2015robust} used an annular uncertainty set to reconstruct the IPNC matrix and constrain interferers, showing similar performance to the beamformer in \cite{gu2012robust}. Nevertheless, reconstructing the IPNC matrix using a complex annular uncertainty set leads to high computational complexity. The work in \cite{zhang2016interference} studied a partial power spectrum sampling method using the covariance matrix taper technique to reconstruct the IPNC matrix with low computational complexity, but this method requires a relatively large number of array elements. The algorithm in \cite{chen2018adaptive} is based on constructing an IPNC matrix directly from the signal-interference subspace. The approach in \cite{mohammadzadeh2018adaptive} utilizes the beamformer output power to jointly estimate the theoretical IPNC matrix and the mismatch by employing eigenvalue decomposition of the received signal covariance matrix. The work in \cite{chen2018robust} proposed an approach based on weighted subspace-fitting for IPNC matrix reconstruction beamformers, specifically designed to mitigate the effect of sensor position errors. A low-complexity beamformer in \cite{mohammadzadeh2019robustcssp} is also presented using the square of the sample covariance matrix in the Capon estimator, estimated based on a correlation sequence.\\
\indent In \cite{zheng2018covariance}, the reconstruction of the IPNC matrix and the estimation of the desired signal are based on a procedure similar to that of \cite{gu2012robust} and \cite{li2003robust}. However, the accuracy of the interference calculation can be influenced by ad-hoc parameters. The method in \cite{Saeed2020} adopts the maximum entropy power spectrum to replace the routine Capon spectrum estimator in the reconstruction process. The beamformer in \cite{zhang2021adaptive} uses a method that separates the SOI component from the training data with a blocking matrix. The SOI steering vector is estimated as the principal eigenvector of the desired signal covariance matrix. Then, the SOI-free data is utilized to calculate the quasi-IPNC matrix. \\
Moreover, the study \cite{gu2012robust} shows that the Capon beamformer delivers strong performance even with errors in the SOI's array SV. However, this analysis did not consider errors in the interference's array SV \cite{somasundaram2014degradation}. Furthermore, the accuracy of the Capon spatial spectrum decreases significantly when coherent signals with linespectra are present \cite{wang2016robust}. Another algorithm, presented in \cite{sun2021robust}, utilizes the gradient vector and IPNC matrix reconstruction by estimating the interference SVs and their powers. Although the aforementioned methods for IPNC matrix reconstruction significantly improve beamforming performance, they require numerical integration with a large number of sampling points, leading to increased beamforming complexity. Additionally, they rely on prior information about the number of interfering sources, DoAs, and the corresponding powers.
\subsection{Contribution}
Motivated by the above-mentioned works, we introduce a novel IPNC matrix reconstruction-based method to enhance beamformer performance by addressing model errors and ensuring robustness against mismatches. The proposed method's core concept is preprocessing-based spatial sampling (PPBSS), which avoids estimating the power and corresponding SV of the interference signals.
Initially, the interfering sources' DoAs are estimated over the available snapshots, and the IPNC matrix is estimated based on a preprocessing matrix. Subsequently, a generalized linear combination of the estimated and identity matrix is utilized to reconstruct the precise IPNC matrix. Specifically, the mean squared error (MSE) between the theoretical and estimated IPNC matrices is employed to achieve a more accurate reconstruction of the IPNC matrix. Following this, we exploit the angular sector of the desired signal to construct the corresponding covariance matrix. The power and SV of the desired signal are then estimated by the eigenvalue and eigenvector lying within the interval of the presumed SOI angular region. To accomplish this, we develop a method based on the power approach \cite{golub1996matrix}, employing a straightforward iterative strategy to compute the dominant eigenvalues and their corresponding eigenvectors.\\
Notably, this method avoids the need for estimating and constructing the noise covariance matrix, thus greatly simplifying the process and enhancing robustness against model errors. Moreover, the main difference between the proposed method and other methods is that we have shown that we can use the preprocessing matrix instead of the sample covariance matrix (SCM) in the shrinkage method. The key contributions and findings of the paper are outlined as follows: 
\begin{itemize}
    \item A proposed algorithm dynamically computes the number of interference sources and their uncertain angular sector per snapshot, enabling real-time adaptation and accurate DoA estimation for time-varying interferences.
    \item We introduce a novel pre-processing covariance matrix based on the computed angular sector, offering a comprehensive representation of the spatial correlation of signals from interference angles.
    \item We utilize the well-known general linear combination algorithm to reconstruct the IPNC matrix, demonstrating that the preprocessing matrix can replace the SCM in the shrinkage method.
    \item We propose an efficient power-method-based algorithm to compute the principal eigenvalue and eigenvector of the desired signal covariance matrix without eigenvalue decomposition (EVD). The eigenvector is then used to estimate the array steering vector.
\end{itemize}

The proposed PPBSS method is evaluated and compared with well-known techniques through numerical analysis. Results show that it achieves a higher SINR across various mismatch scenarios.

The rest of this paper is organized as follows. Section II introduces the basic model and background of the system. Section III presents an algorithm to estimate the DoA of interference. Then, the pre-processing matrix is introduced to reconstruct the IPNC matrix. In section IV, the SV of the desired signal is represented, and the proposed algorithm is detailed. Section V presents the computational complexity of the proposed method versus that of other algorithms and a mathematical analysis of the proposed method with a simple example. Section VI illustrates and discusses the simulation results. Finally, the conclusion is presented in Section VI.

\textit{Notations:}
We use lowercase boldface letters for vectors and uppercase boldface letters for matrices. $\mathbb{E} \{\cdot\}$ and $\Re \{\cdot\}$ stand for the statistical expectation of random variables and the real part of a complex number, respectively. The symbols $\mathbb{C}$ and $\mathbb{R}$ are used to define complex and real numbers, respectively. $(\cdot)^\mathrm{H}$, $ (\cdot)^\mathrm{T} $, and $\text{Tr}(\cdot)$ denote the conjugate transpose, transpose, and trace of a matrix, respectively. The Frobenius norm of a matrix is denoted by $\|\cdot \|$, and $\circ$ denotes the Hadamard product. 

\section{Signal Model and Problem Background}
Consider a linear antenna array of $ L$ sensors with interelement spacing $d$. The data received at the $t$-th  snapshot is described by
  \begin{align}\label{Received Data Vector}
\mathbf{x}(t)= \mathbf{x}_\mathrm{s}(t)+\mathbf{x}_\mathrm{i}(t)+\mathbf{x}_\mathrm{n}(t),
\end{align}
where $\mathbf{x}_\mathrm{s}(t)=s(t)\mathbf{a}(\theta_\mathrm{s})$, $\mathbf{x}_\mathrm{i}(t)=\sum_{p=1}^P i_p(t) \mathbf{a}(\theta_p)$, $\mathbf{x}_\mathrm{n}(t)$ is an independent and identically distributed zero-mean Gaussian noise vector with autocovariance matrix given by $\sigma^2_\mathrm{n} \mathbf{I}$, $P$ is the number of interfering signals, and $\mathbf{I}$ is $L \times L$ identity matrix; $s(t)$, $i_p(t)$, and $\sigma^2_\mathrm{n}$ denote the desired signal, interference signal, and noise variance, respectively. Assume that the desired signal, interference, and noise are statistically independent. The angles $\theta_\mathrm{s}$ and $\theta_p$ denote the DoAs of the desired signal and the $p$-th interference, respectively. The vector $\mathbf{a}(\cdot)$ is the corresponding SV, which has the form $\mathbf{a}(\theta)= (1/\sqrt{L})[1, \ e^{-j
2\pi{d} \sin\theta},  \ \cdots, \ e^{-j
2\pi (L-1){d}\sin\theta}]^{\mathrm{T}}$, where ${d}=\lambda/2$, $\lambda$ is the wavelength, $\theta$ is the DoA. Assuming that the SV $\mathbf{a}(\theta_\mathrm{s}) $ is known, then for a beamformer $\mathbf{w}$, the performance is measured by the output signal-to-interference-plus-noise ratio (SINR)  
\begin{align}\label{SINR}
\mathrm{SINR}= \frac{\sigma^{2}_\mathrm{s} |\mathbf{w}^\mathrm{H} \mathbf{a}(\theta_\mathrm{s})|^2}{ \mathbf{w}^\mathrm{H} \mathbf{R}_\mathrm{i+n}\mathbf{w}},
\end{align}
where $ \sigma^{2}_\mathrm{s} $ is power of the desired signal, $ \mathbf{R}_\mathrm{i+n} $ is the IPNC matrix.Assuming that the interfering signals are independent, the covariance matrix of $\mathbf{x}(t)$ is given by
\begin{align}\label{Theoretical R1}
    \mathbf{R}&=  \sigma^{2}_\mathrm{s}\mathbf{a}(\theta_\mathrm{s})\mathbf{a}^\mathrm{H}(\theta_\mathrm{s})+\sum_{p=1}^P \sigma^{2}_p\mathbf{a}(\theta_p)\mathbf{a}^\mathrm{H}(\theta_p)+\sigma^{2}_\mathrm{n}\mathbf{I},
\end{align}
where $\sigma^2_p$ represents the power of the $p$-th interference component and the theoretical IPNC is defined as
\begin{align} \label{theoretical IPNC}
    \mathbf{R}_\mathrm{i+n} = \sum_{p=1}^P \sigma^{2}_p\mathbf{a}(\theta_p)\mathbf{a}^\mathrm{H}(\theta_p)+\sigma^{2}_\mathrm{n}\mathbf{I}.
\end{align}
Typically, the number of signals, their true SVs, and their power levels are unknown. Therefore, reconstructing the theoretical covariance matrix \( \mathbf{R} \) requires knowledge of the spatial power spectrum $\sigma^2(\theta)$ across all potential directions: 
\begin{align}\label{proposed Ri+n}
\mathbf{R}=\int_{-\pi}^{\pi} \sigma^2(\theta) \mathbf{a}(\theta)\mathbf{a}^\mathrm{H}(\theta)d\theta\approx \sum_{q=1}^{Q}
\sigma^2(\theta_{q})  \mathbf{a}(\theta_{q})\mathbf{a}^\mathrm{H}(\theta_{q}) \Delta \theta_q,
\end{align}
where $\Delta\theta_q \approx \frac{2\pi}{Q}$ and $Q$ is the number of sampling points. It should be noted that $\sigma^2(\theta)$ can be estimated using Capon, entropy, or other types of spatial spectrum estimators \cite{stoica2005spectral}. \\
The problem of maximizing the SINR in (\ref{SINR}) can be cast as:
\begin{align}\label{MVDR}
\underset{{\mathbf{w}}}{\operatorname{min}}\ \mathbf{w}^\mathrm{H} \mathbf{R}_\mathrm{i+n} \ \mathbf{w}\ \hspace{.4cm} \mathbf{s.t.} \hspace{.4cm} \mathbf{w}^\mathrm{H} \mathbf{a}(\theta_\mathrm{s})=1.
\end{align} 
The solution to \eqref{MVDR} is known as the minimum variance distortionless response (MVDR) beamformer and is given by 
\begin{align}\label{optimal weight vector}
\mathbf{w}_{\mathrm{opt}}= \frac{\mathbf{R}_\mathrm{i+n}^{-1} \mathbf{a}(\theta_\mathrm{s})}{\mathbf{a}^\mathrm{H}(\theta_\mathrm{s}) \mathbf{R}_\mathrm{i+n}^{-1}\mathbf{a}(\theta_\mathrm{s})}.
\end{align}
However, $\mathbf{R}_\mathrm{i+n}$ is unavailable in practice, and it is often replaced by the SCM 
\begin{align} \label{Estimated R}
   \hat{\mathbf{R}}= \frac{1}{K}\sum_{t=1}^{K} \mathbf{x}(t)\mathbf{x}^\mathrm{H}(t), 
\end{align}
where  $K$ is the number of snapshots.
\section{The Proposed PPBSS-IPNC Algorithm}
In this section, we propose a new method for reconstructing the IPNC matrix for robust adaptive beamforming (RAB). Our approach involves estimating the DoA of interference, using a preprocessing algorithm to reconstruct the covariance matrix of interference and noise, and then developing a low-complexity method to estimate the true SV of the SOI.

\subsection{Interference DoAs and pre-processed matrix estimation}

\indent Detecting the number of sources hitting a sensor array is crucial. This is essential for super-resolution estimation methods, which often rely on prior knowledge of the number of signals. To eliminate the need for this prior information, we have developed a technique to estimate the DoAs of the interferers from the available snapshots, in which the angular sectors of the interfering signals are computed adaptively.\\
In this regard, we utilize the algorithm in \cite{mohammadzadeh2019robust} where a DoA estimation technique using correlation is introduced, while it is assumed that the interference power is significantly higher than the desired signal power. First, from the set of snapshots, coarse estimates of the DoAs obtained using the discrete Fourier transform (DFT) of the first received vector $\mathbf{x}(1)=[x_1(1),\cdots,x_L(1)]^\mathrm{T}$
where $x_{\ell} (t)$ is the received signal at the \textit{l}-th sensor. Then, an angular sector centered on the estimated DoA is finely scanned, and the angle that maximizes the magnitude of the inner product is taken as the DoA estimate 
\begin{align} \label{correlation estimator}
\hat{\theta}_p(t)=\underset{\theta \in \Theta_\mathrm{x}}{\operatorname{argmax}}\ \arrowvert\mathbf{x}^\mathrm{H}(t)\mathbf{a}(\theta)\arrowvert, \quad t=1,\cdots,K,
\end{align}
where $ \Theta_\mathrm{x}= [\hat{\theta}_p - c, \hat{\theta}_p + c] $ for $p=1, \cdots, P$, is the angular sector corresponding to the estimated interference signal from the DFT process while $c\ll \pi $ is a small angle. Note that the parameter $c$ is chosen based on tuning via simulations and is also supported by the study in \cite{zakharov1999frequency, zakharov2001dft }. Determining the uncertainty region for each interferer involves repeating the same steps for the next snapshot. Once all snapshots have been processed $t=1,2,\cdots, K$, the set of estimated DoA, $[\hat{\theta}_p(1), \hat{\theta}_p(2), \cdots, \hat{\theta}_p(K) ]$ is fitted with a polynomial of a sufficiently high degree (in our proposed method, the degree is assumed to be 2), $\theta_{p_\mathrm{pol}}(t)= \textit{fit}\big([\hat{\theta}_p(1), \hat{\theta}_p(2), \cdots, \hat{\theta}_p(K) ], 'poly2' \big)$. Note that the polynomial determines the angular range in which the interference DoA varies. Defining the angular range in which the interference DoA varies as $\Xi_{ \theta_p}=\max(\theta_{p_\mathrm{pol}}(t))-\min(\theta_{p_\mathrm{pol}}(t)) $ {for $1 \le t \le K$} and sampling the range $[0, 2\pi)$ with $Q$ samples $2\pi q/Q$, $0\le q\le Q-1$, then the width of this range is calculated as follows
\begin{equation} \label{Theta_l}
B_{\theta_p}=\mathrm{ceil} \Big(\dfrac{\Xi _{\theta_p}}{2\pi /Q} \Big), 
\end{equation}
where ceil($\alpha$) truncate each element of $\alpha$ to the nearest integer less than or equal to $\alpha$. We assume the range $(g_p\leq g \leq g_p+{B_{\theta_p}}-1)$ corresponds to an angular sector centered on the DoA of the $p$-th interference. The index of the first angular interval of this range, $g_p$ can be calculated as
\begin{equation} \label{ni}
g_p = \mathrm{ceil} \Big( N_{pc}-\dfrac{B_{\theta_p}}{2} \Big),
\end{equation}
where $  N_{pc}=\mathrm{ceil}(\frac{\theta_{pc}}{2\pi /Q}) $ is the number of angular intervals corresponding to the center of the range assumed to be $ \theta_{pc}=\big(\max(\theta_{p_\mathrm{pol}}(t))+\min(\theta_{p_\mathrm{pol}}(t) ) \big)/2 $. 

Next, we show that the IPNC matrix can be accurately constructed based on an estimate of the DoAs of the interference without directly resorting to the interference SVs' power spectrum. We employ a preprocessing matrix that works on an approximation of the orthogonal space of the presumed SOI steering vector. This approach uses a subspace procedure to achieve an effective approximation for RAB purposes. We define the matrix $\mathbf{C}$ as a preprocessing IPNC matrix as follows
\begin{align} \label{C_L}
    \mathbf{C}=  \sum _{p=1}^{P} \sum _{\ell = g_p} ^{g_p+{B_{\theta_p}}-1} \mathbf{a}(\psi_\ell) \mathbf{a}^\mathrm{H}(\psi_\ell),
\end{align}
where $\psi_{\ell} = \frac{2 \pi}{Q}(\ell-1)$ and $g_p \le \ell \le g_p+{B_{\theta_p}}-1 $ is the range corresponds to a set of discrete angles around the DoA of the $p$-th interference. It is evident that this matrix provides a comprehensive measure of the spatial correlation of signals arriving from the set of interference angles. 
Since each term $\mathbf{a}(\psi_\ell) \mathbf{a}^\mathrm{H}(\psi_\ell)$ represents a rank-1 projection matrix corresponding to a specific interference angle $\psi_{\ell}$, summing over all interference angles $\psi_{\ell}$ within the range $[g_p, g_p+{B_{\theta_p}}-1]$ captures the spatial correlation of all signals within that angular range corresponding to the $p$-th interference.

\subsection{IPNC Matrix Reconstruction}
In order to employ the matrix $\mathbf{C}$ to help with the estimation of the IPNC matrix, we can utilize the general-linear-combination-based covariance matrix estimation, which is based on a shrinkage method \cite{ledoit2004well}. We consider a combination of the preprocessed covariance matrix $\mathbf{C}$ and the identity matrix $\mathbf{I}$ to obtain preferably a more accurate estimate of the IPNC matrix ${\mathbf{R}}_\mathrm{i+n}$ as follows,
\begin{align} \label{tilde Ri+n}
    \tilde{\mathbf{R}}_\mathrm{i+n} = \eta \mathbf{I} + \rho \mathbf{C}.
\end{align}
Let $\eta_0$ and $\rho_0$ denote optimal values of $\eta$ and $\rho$ that make \eqref{tilde Ri+n} a good estimate of the theoretical matrix in \eqref{theoretical IPNC}. We aim to obtain estimates $\hat{\eta}_0$ and $\hat{\rho}_0$ of $\eta_0$ and $\rho_0$ from available received data and preprocessed matrix $\mathbf{C}$.
Here, the shrinkage parameters $\eta \geq 0$ and $\rho \geq 0$ are chosen by minimizing the mean squared error (MSE): $\text{MSE} = \mathbb{E} \{ \| \tilde{\mathbf{R}}_\mathrm{i+n} - \mathbf{R}_\mathrm{i+n} \|^2 \}$. By substituting \eqref{tilde Ri+n} into MSE, we can write (please refer to Appendix A for the proof)
\begin{align} \label{MSE}
\text{MSE}   =& \mathbb{E} \left\{ \|\eta \mathbf{I}-(1-\rho) \mathbf{R}_\mathrm{i+n}+ \rho (\mathbf{C}-\mathbf{R}_\mathrm{i+n}) \|^2 \right\} \nonumber \\
 =& \|\eta \mathbf{I}-(1-\rho) \mathbf{R}_\mathrm{i+n} \|^2+\rho^2 \mathbb{E} \left\{ \|\mathbf{C}-\mathbf{R}_\mathrm{i+n} \|^2 \right\} \nonumber \\
 =& \eta^2 L-2 \eta(1-\rho) \operatorname{Tr}(\mathbf{R}_\mathrm{i+n}) \nonumber \\
 +&(1-\rho)^2 \| \mathbf{R}_\mathrm{i+n} \|^2+\rho^2 \mathbb{E} \left\{\|\mathbf{C}-\mathbf{R}_\mathrm{i+n} \|^2\right\}.
\end{align}
 The unconstrained minimization of \eqref{MSE} with respect to $\eta$ for fixed $\rho$, gives
\begin{align} \label{alpha_0}
    \eta_0 = \frac{(1-\rho_0) \text{Tr}(\mathbf{R}_\mathrm{i+n})}{L},
\end{align}
where $\rho_0$ is the minimizer of the function that is obtained by inserting \eqref{alpha_0} in \eqref{MSE},
\begin{equation} \label{frac}
    (\rho_0-1)^2  \Big [\| \mathbf{R}_\mathrm{i+n} \|^2 - \frac{\text{Tr}^2(\mathbf{R}_\mathrm{i+n})}{L} \Big ] + \rho^2 \mathbb{E} \{ \| \mathbf{C} - \mathbf{R}_\mathrm{i+n} \|^2 \}.
\end{equation}
The unconstrained minimization of \eqref{frac} with respect to $\rho$ results in
\begin{align} \label{rh0}
    \rho_0 = \frac{\beta}{\beta + \zeta},
\end{align}
where $\zeta \triangleq \mathbb{E} \{ \| \mathbf{C}-\mathbf{R}_\mathrm{i+n} \|^2 \}$ and 
\begin{align}
    \beta = \| \mathbf{R}_\mathrm{i+n} \|^2 -\frac{  \text{Tr}^2(\mathbf{R}_\mathrm{i+n})}{L}.
\end{align}
Based on the Cauchy-Schwarz inequality, it can be seen that $\beta > 0$. This implies that $\rho_0 \in (0,1)$. This condition, along with the fact that $\text{Tr}(\mathbf{R}_\mathrm{i+n}) > 0 $ results in $\eta_0 > 0 $. Thus, $\eta_0$ and $\rho_0$ are constrained minimizers of the MSE.\\ 
Let 
\begin{align} \label{mu}
    \mu =  \frac{\text{Tr}(\mathbf{\mathbf{R}_\mathrm{i+n}})}{L},
\end{align}
we can write that 
\begin{align} \label{beta}
    \| \mu \mathbf{I} - \mathbf{\mathbf{R}_\mathrm{i+n}} \|^2 = \frac{\text{Tr}^2(\mathbf{R}_\mathrm{i+n})}{L} + \| \mathbf{R}_\mathrm{i+n} \|^2 - 2 \frac{\text{Tr}^2(\mathbf{R}_\mathrm{i+n})}{L} =  \beta
\end{align}
Using \eqref{mu}, $\eta_0$ and $\rho_0$ can be estimated as:
\begin{align} \label{eta rho}
    \hat{\rho}_0 = 1- \frac{\hat{\zeta}}{\hat{\beta} + \hat{\zeta}} , \quad
    \hat{\eta}_0 = (1- \hat{\rho}_0) \hat{\mu}  = \frac{\hat{\zeta}}{\hat{\beta} + \hat{\zeta}} \hat{\mu},
\end{align}
where $\hat{\beta}$, $\hat{\zeta}$ and $\hat{\mu}$ are estimates of $\beta$, $\zeta$ and $\mu$, respectively.
We realize from $\rho_0$ in \eqref{rh0} and $\mu$ in \eqref{mu} that the estimates in \eqref{eta rho} depend on the true interference-plus-noise covariance matrix $\mathbf{R}_\mathrm{i+n}$. However, it is not available in practical scenarios, and we replace it with the sample covariance matrix, $\hat{\mathbf{R}}$. So we can estimate $\mu$ as
\begin{align}
    \hat{\mu}= \frac{\text{Tr}(\hat{\mathbf{R}})}{L}, 
\end{align}
and $\hat{\zeta}$ as
\begin{align} \label{zeta0}
    \hat{\zeta} = \frac{1}{K^2} \sum_{t=1}^K \| \mathbf{x}(t) \|^4 - \frac{1}{K} \| \hat{\mathbf{R}} \|^2.
\end{align}
The proof of \eqref{zeta0} is given in Appendix B. In addition, from the denominator of \eqref{eta rho}, we have 
\begin{align}
    \beta + \zeta &= \| \mu \mathbf{I} - \mathbf{\mathbf{R}_\mathrm{i+n}} \|^2 + \mathbb{E} \{ \| \mathbf{C}-\mathbf{R}_\mathrm{i+n} \|^2 \} \nonumber \\ &= \mathbb{E} \{ \| \mathbf{C}- \mu \mathbf{I} \|^2 \},
\end{align}
an estimate of which is given by $\| \mathbf{C}- \hat{\mu} \mathbf{I} \|^2$. Consequently, we write
\begin{align} \label{rho and eta}
    \hat{\eta}_0 = \frac{\hat{\zeta}}{\| \mathbf{C}- \hat{\mu} \mathbf{I} \|^2} \hat{\mu}, \quad 
    \hat{\rho}_0 = 1- \frac{\hat{\zeta}}{\| \mathbf{C}- \hat{\mu} \mathbf{I} \|^2}.
\end{align}
However, in ( \textbf{Lemma 3.4},\cite{ledoit2004well}), it is proved that the estimate of $\hat{\rho}_0$ in \eqref{rho and eta} may not necessarily be positive.
To ensure non-negative estimates of $\hat{\rho}_0$, we can use modified estimates of $\eta_0$ and $\rho_0$,  
\begin{align} \label{rho and eta final}
    \tilde{\eta}_0=\text{min}(\hat{\eta}_0, \hat{\mu}), \quad 
    \tilde{\rho}_0 = 1- \frac{\tilde{\eta}_0}{\hat{\mu}}.
\end{align}
Then, the reconstructed PPBSS matrix can be written as
\begin{align} \label{Est IPNC}
    \tilde{\mathbf{R}}_\mathrm{i+n} = \tilde{\eta}_0 \mathbf{I} + \tilde{\rho}_0 \mathbf{C}.
\end{align}
\begin{figure}[h]
	\centering
	\includegraphics[height=1.8in]{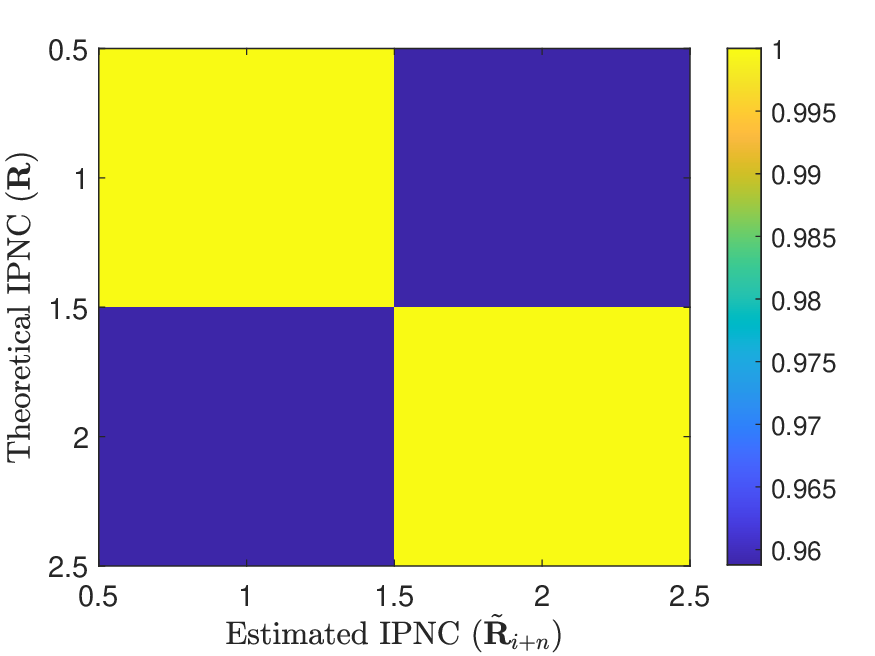}
\vspace{-0.05em}
	\caption{Correlation of theoretical IPNC \eqref{theoretical IPNC} versus proposed IPNC \eqref{tilde Ri+n} for SNR =-20 dB, }
	\label{Correlation}
\end{figure}
To show the validation of the proposed IPNC matrix, we demonstrate the Pearson correlation (Appendix C) between the theoretical IPNC matrix in \eqref{Theoretical R1} and the estimated one based on the proposed approximation in \eqref{Est IPNC}, which is a quantitative measure of the linear association between two matrices and helps to assess their similarity. 

\indent The visual representation of the correlation between two matrices provides a clear insight into their relationship in Fig.~\ref{Correlation}. We observe that the main diagonal contains a correlation coefficient of 1, indicating a perfect correlation between each column of the matrices. This is expected, as each column perfectly correlates with itself. Values on the off-diagonal elements are very close to 1. This suggests a strong positive linear relationship between the corresponding columns of the two matrices. 

\section{Desired signal steering vector estimation}

\indent In order to estimate the SV of the desired signal, we need to reconstruct the desired signal covariance matrix. Recalling \cite{Saeed2020}, we propose an approach that leverages the power spectrum estimate across all possible directions, and coarse estimates of the angular regions where the SOI lies as described by 
\begin{align}\label{proposed Ri+n}
\hat{\mathbf{R}}_{\mathrm{s}}=\int_{{\Theta}_\mathrm{s}} \hat{\sigma}^2(\theta) \mathbf{a}(\theta)\mathbf{a}^\mathrm{H}(\theta)d\theta,
\end{align}
where $\hat{\sigma}^2(\theta)=1 \big/{\alpha  \vert  \mathbf{a}^\mathrm{H}(\theta)\hat{\mathbf{R}}^{-1} \mathbf{d}_1 \vert ^2}$
is the maximum entropy power spectrum estimate, and $\Theta_\mathrm{s}$ is the angular sector of the SOI region, which can be defined based on low-resolution finding methods \cite{capon1969high,stoica1990performance}; Also, $\mathbf{d}_1=[\begin{smallmatrix}1, & 0, & \cdots, & 0\end{smallmatrix}]^\mathrm{T}$,   $\alpha=1/\mathbf{d}_1^\mathrm{T}\hat{\mathbf{R}}^{-1}\mathbf{d}_1$.
Sampling $\Theta_\mathrm{s}$ uniformly with $S \leq L$ (since $\Theta_\mathrm{s}$ is usually a small sector) sampling points 
spaced by $\Delta\theta_{s}$, (\ref{proposed Ri+n}) can be approximated by
\begin{align} \label{Summation}
\hat{\mathbf{R}}_{\mathrm{s}} \approx \sum_{s=1}^{S} \hat{\sigma}^2(\theta_{s}) \mathbf{a}(\theta_{s})\mathbf{a}^\mathrm{H}(\theta_{s})
 \Delta\theta_{s},
\end{align}
where $\mathbf{a}(\theta_{s})$ is the SV associated with $\theta_{s}$ belonging to the discrete angular sector $\Theta_{s}$. Since only one desired signal is assumed in the uncertainty region ($\Theta_\mathrm{s}$), the principal eigenvector of the constructed matrix will be the SV we are looking for. To avoid the EVD and reduce the complexity, the power method in \cite{golub1996matrix,tammen2018complexity} is employed to estimate the principal eigenvalue and the corresponding eigenvector of the reconstructed desired signal covariance matrix, $\hat{\mathbf{R}}_\mathrm{s}$, utilizing  Algorithm 1. {Employing the power method in place of a full eigenvalue decomposition reduces computational complexity from $\mathcal{O}(L^3)$ to $\mathcal{O}(L^2)$, which is particularly beneficial for real-time applications of the diffuse power spectral density estimator, especially in scenarios involving a large number of array elements $L$.}

\begin{figure}[h]
	\centering
	\includegraphics[height=2.1in]{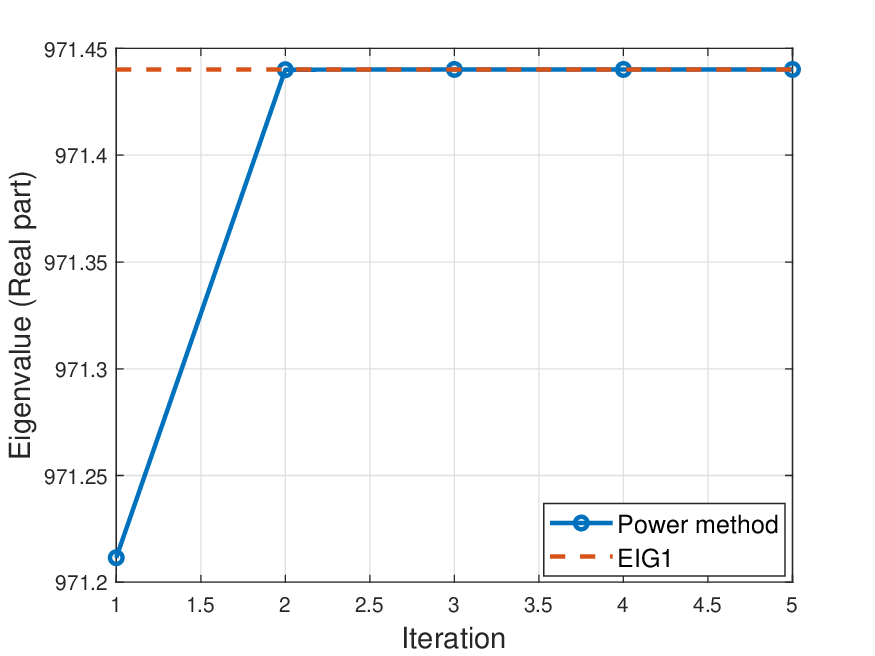}
\vspace{-0.05em}
	\caption{Eigenvalue calculation of the power method versus the iterations }
	\label{Eigen}
\end{figure}

In this Algorithm, the principal eigenvalue $\kappa =\hat{\sigma}_\mathrm{s}^2$ and the eigenvector $\mathbf{b}=\mathbf{a}(\hat{\theta}_\mathrm{s})$ of $\hat{\mathbf{R}}_\mathrm{s}$ can be computed in only two iterations. {Fig.~\ref{Eigen} illustrates the comparison between the principal eigenvalue obtained using the full eigenvalue decomposition (denoted as EIG1) and the result produced by the iterative power method for the input SNR = 10 dB. As shown in the figure, the power method rapidly converges to the dominant eigenvalue computed via full EVD, achieving near-identical results within just two iterations. This demonstrates the efficiency and accuracy of the power method in approximating the leading eigenvalue with significantly reduced computational overhead.}

The robust beamformer for the SOI direction $\hat{\theta}_\text{s}$ is then computed by
\begin{align}\label{proposed weight vector}
\mathbf{w}_{\mathrm{PPBSS}}=\dfrac{\tilde{\mathbf{R}}_\mathrm{i+n}^{-1} \mathbf{a}(\hat{\theta}_\mathrm{s})}{\mathbf{a}^\mathrm{H}(\hat{\theta}_\mathrm{s}) \tilde{\mathbf{R}}_\mathrm{i+n}^{-1}\mathbf{a}(\hat{\theta}_\mathrm{s})}.
\end{align}

\begin{algorithm}[!]
\caption{Proposed PPBSS Algorithm}
\begin{algorithmic}[1]
\STATE \textbf{Input:}\: $\mathbf{b}_0 \in \mathbb{C}^M$, $\mathbf{d}_1=[\begin{smallmatrix}1, & 0, & \cdots, & 0\end{smallmatrix}]^\mathrm{T}$, $\text{Maximum iteration} = \textit{Iter}_\text{max}$,  $\delta = 0.001$, Array received data vector$\lbrace \mathbf{{x}}(k) \rbrace_{k=1}^K $;
\STATE \textbf{Initialize:}\: Compute the sample covariance matrix
\STATE Compute the DoA of interference based on \eqref{correlation estimator};
\STATE Calculate $B_{\theta_p}$ and $g_p$ from \eqref{Theta_l} and \eqref{ni};
\STATE Define $\Psi=\bigcup \limits_{p=1}^{P}\{g_p\leq g \leq g_p+{B_{\theta_p}}-1 \}$;
\STATE Estimate the matrix $\mathbf{C}$, \eqref{C_L};
\STATE Calculate $\eta_0$ and $\rho_0$ using \eqref{rho and eta final};
\STATE Construct SOI matrix, $\hat{\mathbf{R}}_\mathrm{s}$, according to \eqref{Summation};
\STATE \quad \textbf{For} \ $ j=1 $ \textbf{to}  $\textit{Iter}_{max}$ \textbf{do}; 
\STATE \qquad  $\mathbf{v}_j=\hat{\mathbf{R}}_\mathrm{s} \mathbf{b}_{j-1}$ ,  $\mathbf{b}_{j}=\mathbf{v}_j / \|\mathbf{v}_j\|_2 $
\STATE \qquad Define \  $err=\sqrt(1-(\mathbf{b}_j^\mathrm{H} \mathbf{b}_{j-1})^2) $;
\STATE \qquad \ \textbf{If} \quad $err < \delta$ ;\\
\qquad \qquad $\mathbf{b}=\mathbf{b}_j$;\\
\qquad \qquad \textbf{break};\\
\quad \quad \textbf{else} \quad  $\mathbf{b}_{j-1}=\mathbf{b}_j$;\\
\qquad \textbf{end If}\quad
\STATE \quad \textbf{End}
\STATE  $\kappa=\mathbf{b}_{j}^\mathrm{H} \hat{\mathbf{R}}_\mathrm{s} \mathbf{b}_{j}$;
\STATE $ \mathbf{a}(\hat{\theta}_\mathrm{s}) = \mathbf{b}$;
\STATE \textbf{Output:}\: Proposed beamformer, $\mathbf{w}_\text{PPBSS}$ as in \eqref{proposed weight vector};
\end{algorithmic}
\end{algorithm}

\section{Analysis of PPBSS}
In this section, we analyze {the DoA estimation accuracy,} the array beampattern of the proposed PPBSS RAB technique, and assess its computational complexity compared to competing RAB approaches.

{\subsection{Performance Analysis of DoA Estimation Using CRB}}
{To evaluate the accuracy of the proposed DoA estimation method, we consider the Cramér-Rao Bound (CRB) as a theoretical benchmark that characterizes the minimum achievable estimation error given in \cite{stoica2002music} as follows}
{
\begin{align}
\mathrm{CRB}_\theta^{\mathrm{C}}=\frac{\sigma^2_\mathrm{n}}{2 K}\left\{\Re\left[\mathbf{H} \circ \hat{\mathbf{R}}^T\right]\right\}^{-1},
\end{align}
where $\hat{\mathbf{R}}$ is defined in \eqref{Estimated R}, and}

{\begin{align}
& \mathbf{H}=\dot{\mathbf{A}}^\mathrm{H}\left[\mathbf{I}-\mathbf{A}\left(\mathbf{A}^\mathrm{H} \mathbf{A}\right)^{-1} \mathbf{A}^\mathrm{H}\right] \dot{\mathbf{A}} \\
& \mathbf{A} = [\mathbf{a}(\theta_\mathrm{s}), \  \mathbf{a}(\theta_1) \cdots \mathbf{a}(\theta_p) \cdots \mathbf{a}(\theta_P)]\\
& \dot{\mathbf{A}}=\left[\begin{array}{llll}
\frac{\partial \mathbf{a}\left(\theta_s\right)}{\partial \theta_s}, & \frac{\partial \mathbf{a}\left(\theta_1\right)}{\partial \theta_1} & \cdots & \frac{\partial \mathbf{a}\left(\theta_{\mathrm{P}}\right)}{\partial \theta_P}
\end{array}\right] 
\end{align}}

\begin{figure}[h]
	\centering
	\includegraphics[height=2.5in]{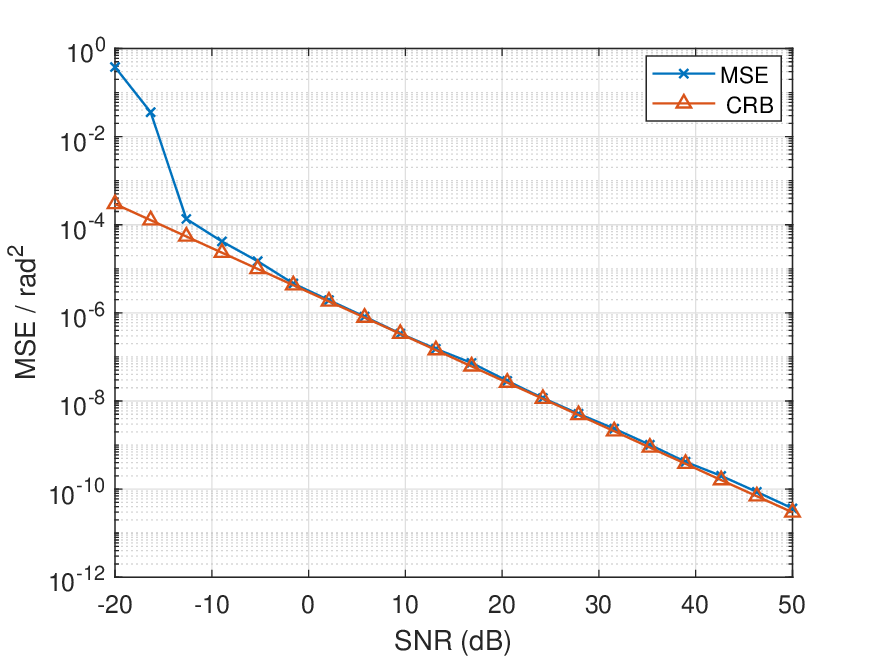}
	\caption{Performance evaluation of the proposed PPBSS method under the interference DoA estimator. }
	\label{CRB}
\end{figure}

{Fig.~\ref{CRB} illustrates the MSE of the proposed DoA estimation method compared with the CRB as a function of the SNR. As expected, the CRB decreases monotonically with increasing SNR, indicating the theoretical lower bound on the estimation variance. The MSE closely follows the CRB for moderate to high SNRs (above 0 dB), demonstrating that the proposed estimator approaches the theoretical efficiency limit in these regimes. However, at low SNRs (below –10 dB), the MSE deviates noticeably from the CRB due to increased noise and limited information content, which is typical for most practical estimators. Overall, the figure confirms that the proposed method is near-optimal at high SNRs and performs robustly across a wide SNR range.}

\subsection{Computational Complexity}
The computational complexity of the PPBSS is compared with that of some recently proposed methods. Given an array of $L$ elements, $Q$ angles inside the interference region, and $S$ sampling points of the desired signal angular sector, the computational complexity for computing the IPNC matrix \eqref{tilde Ri+n} is $ \mathbb{O}(KL+L^2)$ and $ \mathbb{O}(L^2S)$ to calculate the desired signal matrix. Therefore, the computational complexity of the proposed PPBSS algorithm is $ \mathbb{O}(\mathrm{max}(KL+L^2, L^2S))$. Meanwhile, the performance of the beamformers in \cite{zheng2018covariance,luo2023urglq} is dominated by solving a QCQP technique, which has $ \mathbb{O}(L^{3.5})$ complexity. The computational complexity of the algorithm in \cite{huang2012modified} is $\mathbb{O}(\max (L^{3.5}, QL^2)) $. The beamformer in \cite{Saeed2020} needs $ \mathbb{O}(QL^2)$ and $ \mathbb{O}(SL^2)$ complexity for IPNC matrix reconstruction and the desired signal SV estimation, where $Q$ is the number of uniform samples in the interference-plus-noise region. The reconstructed IPNC matrix in \cite{zhang2015interference} has a complexity of $\mathbb{O}(L^3) $.

\subsection{The Array Beampattern of the Proposed PPBSS}

\begin{figure}[h]
\centerline{\includegraphics[width=0.47\textwidth]{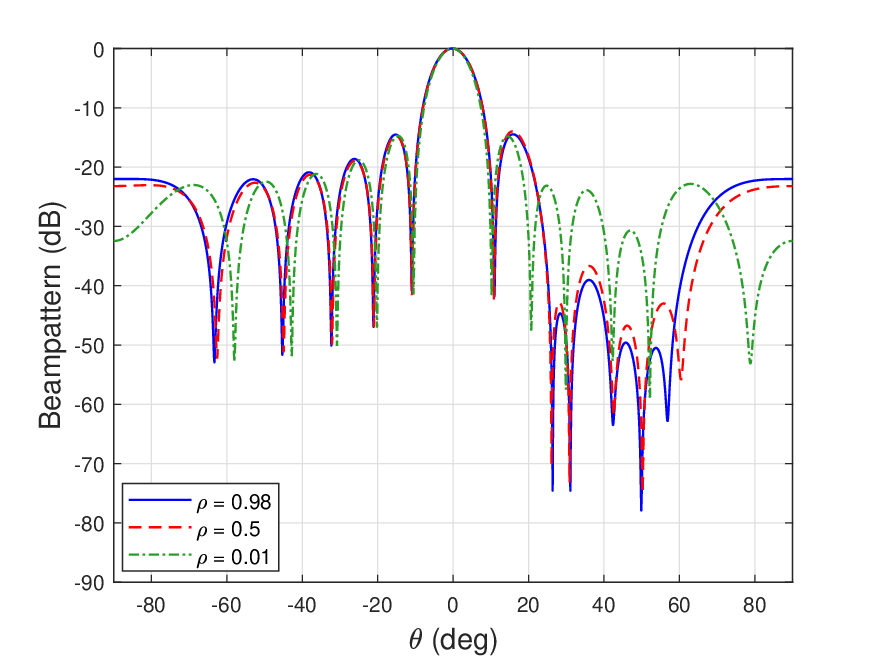}}
\vspace{-0.07em}
\caption{The beampattern for different values of $\rho$, DoA of desired signal=$0^o$ and the DoA of interferences= $30^o, 50^o$}
\label{rho}
\end{figure}

\begin{figure}[h]
\centerline{\includegraphics[width=0.47\textwidth]{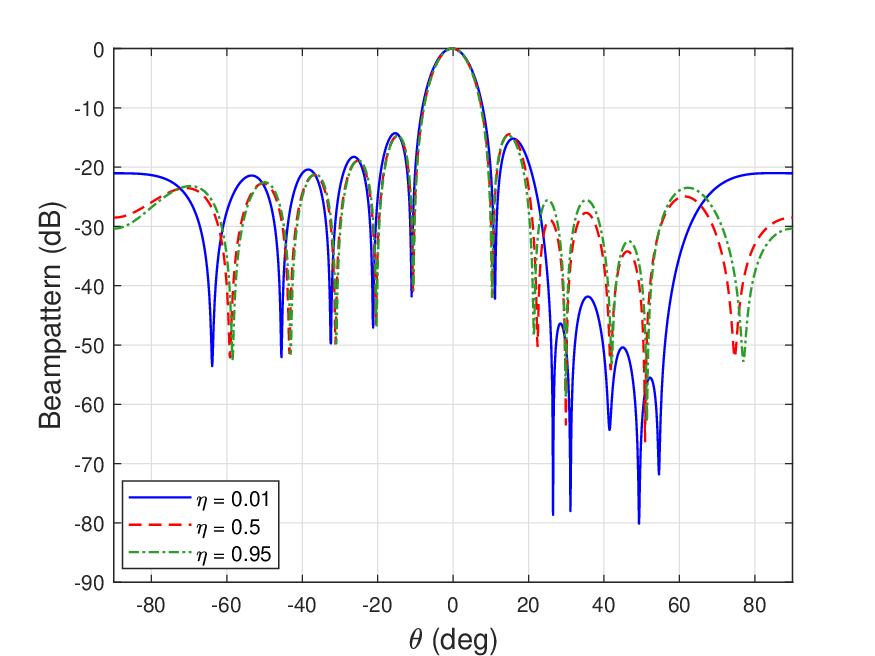}}
\vspace{-0.07em}
\caption{The beampattern for different values of $\eta$, DoA of desired signal=$0^o$ and the DoA of interferences= $30^o, 50^o$}
\label{eta}
\end{figure}

In this section, we will derive the directional response of the array in the angular sector of interferences to evaluate the effectiveness of the proposed PPBSS technique. To accomplish this, we will use the example of a single interference by examining the relationship between the depth of the notch in the beampattern and the weight vector. This analysis will be useful in understanding the performance of the proposed approach.\\
Let the EVD of $ \mathbf{C}$ be written as 
\begin{align}\label{R_r}
\mathbf{C} =\mathbf{E} \mathbf{\Gamma} \mathbf{E}^\mathrm{H}=\sum_{r=1}^{R^a} \gamma_r \mathbf{e}_r \mathbf{e}^\mathrm{H}_r,
\end{align}
where $\mathbf{\Gamma} $ is a diagonal matrix with the eigenvalues $\{ \gamma_r \}_{r=1}^{R^a}$ of $\mathbf{C}$  on the diagonal, and $\mathbf{E}$ is an orthogonal matrix with a corresponding set of orthonormal eigenvectors $\mathbf{e}_1, \cdots, \mathbf{e}_{R^a}$ as columns and $ (\text{rank}(\mathbf{C})={R^a} \leq L) $ denotes the rank of $ \mathbf{C} $. The rank $R^a$ depends on the width of the angular sector $ B_{\theta_p} $ and is one if it shrinks to zero. However, the dominant eigenvalue would be of the order of $ \gamma_\text{max}\approx L|B_{\theta_p}| $ and the majority of the eigenvalues would be close to zero for a sufficiently small width or $|B_{\theta_p}| \to 0$. By substituting (\ref{R_r}) back into \eqref{tilde Ri+n} and taking the inverse of $ \tilde{\mathbf{R}}_\mathrm{i+n} $ based on the Woodbury matrix inversion lemma, we can show that
\begin{align}\label{C inv}
\tilde{\mathbf{R}}_\mathrm{i+n}^{-1}=\frac{1}{\eta}\Big[\mathbf{{I}}_M-\mathbf{E}\Big( \frac{\eta}{ \rho} \mathbf{\Gamma}^{-1}+\mathbf{{E}}^\mathrm{H}\mathbf{{E}} \Big)^{-1} \mathbf{{E}}^\mathrm{H} \Big].
\end{align}
The beam response is a tool used to analyze the performance of a beamformer by examining the response for a proposed weight vector, denoted as $\mathbf{w}_{\mathrm{PPBSS}}$, as a function of angle $\theta$. This angular response is computed by applying the beamformer, denoted as $\mathbf{w}_{\mathrm{PPBSS}}$, to a set of array response vectors from all possible angles ranging from $- \pi/2$ to $ \pi/2$:
\begin{align}\label{beampattern}
\mathbf{D}(\theta)=\big \rvert \mathbf{w}_{\mathrm{PPBSS}}^\mathrm{H} \mathbf{a}(\theta) \big \rvert = \Big \rvert \dfrac{\mathbf{a}^\mathrm{H}(\theta)  \tilde{\mathbf{R}}_\mathrm{i+n}^{-1} \mathbf{a}(\hat{\theta}_\mathrm{s})}{\mathbf{a}(\hat{\theta}_\mathrm{s})^\mathrm{H}  \tilde{\mathbf{R}}_\mathrm{i+n}^{-1} \mathbf{a}(\hat{\theta}_\mathrm{s})} \Big \rvert.
\end{align}
By substituting \eqref{C inv} in \eqref{beampattern}, the numerator is calculated as
\begin{align}\label{num}
\Big \rvert \mathbf{a}^\mathrm{H}(\theta) \tilde{\mathbf{R}}_{\mathrm{i+n}}^{-1} & \mathbf{a}(\hat{\theta}_\mathrm{s})  \Big \rvert=\dfrac{1}{\eta}\Big\rvert\mathbf{a}^\mathrm{H}(\theta) \mathbf{a}(\hat{\theta}_\mathrm{s}) \nonumber \\
&-\mathbf{a}^\mathrm{H}(\theta)\mathbf{E}\Big( \frac{\eta}{ \rho} \mathbf{\Gamma}^{-1}+\mathbf{{E}}^\mathrm{H}\mathbf{{E}} \Big)^{-1} \mathbf{E}^\mathrm{H} \mathbf{a}(\hat{\theta}_\mathrm{s}) \Big\rvert.
\end{align}
Let define a vector $\mathbf{u}$, for $\theta \in B_{\theta_p}$ which is the orthogonal projection of $ \mathbf{a}(\hat{\theta}_\mathrm{s}) $ onto the subspace spanned by the eigenvectors $ \mathbf{e}_r$ as follows
\begin{align}
\mathbf{{E}}^\mathrm{H} \mathbf{a}(\hat{\theta}_\mathrm{s})=\mathbf{u}=[u_{1}, \cdots, u_{R^a}]^\mathrm{T}. 
\end{align}
Moreover, we will employ the fact that $ \mathbf{E}^\mathrm{H}\mathbf{E}=\mathbf{\mathbf{I}}_{R^a} $ since $\mathbf{E}$ is an orthogonal matrix with a corresponding set of orthonormal eigenvectors $\mathbf{e}_1, \cdots, \mathbf{e}_{R^a}$ of $\mathbf{C}$ as columns. That means that any SV whose DoA comes from $B_{\theta_p}$ can be expressed as a linear combination of the columns of $\mathbf{E}$ \cite{lie2011adaptive}. Therefore, the SV $ \mathbf{a}(\theta)\in\text{span}(\mathbf{E})$ can be expressed as $ \mathbf{{a}}(\theta)=\mathbf{E} \mathbf{h}(\theta) $ for some $\mathbf{h}(\theta)$ ($\theta\in B_{\theta_p} $) {\cite{zhang2013robust}}. Then, \eqref{num} can be written as
\begin{align}
\Big \rvert \mathbf{a}^\mathrm{H}(\theta) \tilde{\mathbf{R}}_{\mathrm{i+n}}^{-1}  \mathbf{a}(\hat{\theta}_\mathrm{s})  \Big \rvert &=\Big\rvert \dfrac{  \mathbf{h}^\mathrm{H}(\theta) }{\eta}  \Big( \mathbf{u} 
-\Big( \frac{\eta}{ \rho} \mathbf{\Gamma}^{-1}+\mathbf{I}_{R^a}\Big)^{-1} {\mathbf{u}} \Big) \Big\rvert \nonumber\\
&=\Big\rvert \sum_{r=1}^{R^a}h^*_r(\theta) u_r \Big(\dfrac{1}{\eta + \gamma_r \rho}\Big) \Big\rvert.
\end{align}
The denominator of equation (\ref{beampattern}) can be expressed as
\begin{align}\label{den}
\Big\rvert \mathbf{a}(\hat{\theta}_\mathrm{s})^\mathrm{H} \hat{\mathbf{R}}_{\mathrm{i+n}}^{-1} \mathbf{a}(\hat{\theta}_\mathrm{s}) \Big \rvert = \dfrac{1}{\eta} \Big[\| \mathbf{a}(\hat{\theta}_\mathrm{s}) \| ^2  - \sum_{r=1}^{R^a} \vert u_r \vert^2 \Big(\dfrac{\gamma_r \rho}{\eta+\gamma_r \rho}\Big) \Big].
\end{align}
If the SOI angular direction is sufficiently separated from the sector$ B_{\theta_p} $, then $\| \mathbf{u} \|_2 \ll \|\mathbf{a}(\hat{\theta}_\mathrm{s}) \|_2$. Moreover, for a sufficiently small sector $ B_{\theta_p} $, the dominant eigenvalue would be much bigger than most of the eigenvalues $ \gamma_r $ of $ \mathbf{C} $, which would be either very small or almost zero. According to these points, the summation in (\ref{den}) can be disregarded, and the beampattern is expressed as
\begin{align} \label{last beampattern}
\mathbf{D}(\theta)=\dfrac{\eta}{\| \ \mathbf{a}(\hat{\theta}_\mathrm{s}) \| ^2 } \Big\vert \sum_{r=1}^{R^a}h^*_r(\theta) u_r \Big(\dfrac{1}{ \eta+\gamma_r \rho}\Big) \Big\vert.  
\end{align}
Fig.~\ref{rho} and Fig.~\ref{eta} visually illustrate the beampattern behavior and effect of \(\rho\) under various values for the parameters defined in the simulation results section, while SNR is fixed at -10 dB. Fig.~\ref{rho} shows that the design becomes more focused on suppressing interference for a larger value of \(\rho\). This results in deeper nulls in the beampattern, effectively reducing the impact of interference signals. In other words, the system prioritizes nullifying the interference signals more aggressively, ensuring they have minimal influence on the SOI. Fig.~\ref{eta} demonstrates that when \(\eta\) is increased, the focus shifts towards the main lobe of the beampattern, potentially at the expense of accuracy in targeting the interference directions. This means that the system becomes less precise in suppressing interference, and the system may "miss" or misalign the nulls intended for interference suppression, allowing some interference to persist.

\section{Simulation Results} \label{simu}
It is assumed that we have a uniform linear array with $L=12$ omnidirectional sensors. There is one desired signal arriving at the array from the presumed direction of ${\theta}_\mathrm{s} = 0^\circ$, while the uncorrelated interference signals come from angles $50^\circ$ and $30^\circ$. The size of the number of snapshots is $K=100$. The additive noise is assumed to be spatially white Gaussian with unit variance. For the sake of brevity, we gave a prefix titled covariance matrix construction (CMC) to the proposed methods and other methods compared in the simulation results. The proposed CMC-PPBSS method is compared with the reconstruction-estimation-based beamformers in \cite{zheng2018covariance} (CMC-EST), {\cite{luo2023urglq} (CMC-URGLQ)}, \cite{khabbazibasmenj2012robust} (CMC-SV), \cite{huang2012modified} (CMC-SUB), \cite{zhang2015interference} (CMC-SPSS), and \cite{Saeed2020} (CMC-MEPS). The two interferers' interference-to-noise ratios are set to 30 dB. In the CMC-EST, $Q$ is set to 200. In the CMC-EST beamformer, the norm of the SV mismatch is constrained to an upper bound of \( \sqrt{0.1} \). Also, the parameter $\rho=0.7$  is considered in CMC-SUB beamformers. For CMC-PPBSS, we employ $P=2$ and $S=12$ samples and perform 100 Monte Carlo runs. The angular sector of the desired signal is set to be $ {\Theta}_\mathrm{s}=[{\hat{\theta}_\mathrm{s}}-4^\circ,{\hat{\theta}_\mathrm{s}}+4^\circ] $ where the interference angular sector is $ \hat{\Theta}=[-90^\circ,{\hat{\theta}_\mathrm{s}}-4^\circ)\cup({\hat{\theta}_\mathrm{s}}+4^\circ,90^\circ] $. The MATLAB CVX toolbox \cite{grant2008cvx} is used to solve all convex optimization problems.

\subsection{Mismatch due to look direction error}
\begin{figure}[t]
	\centering
	\includegraphics[height=2.5in]{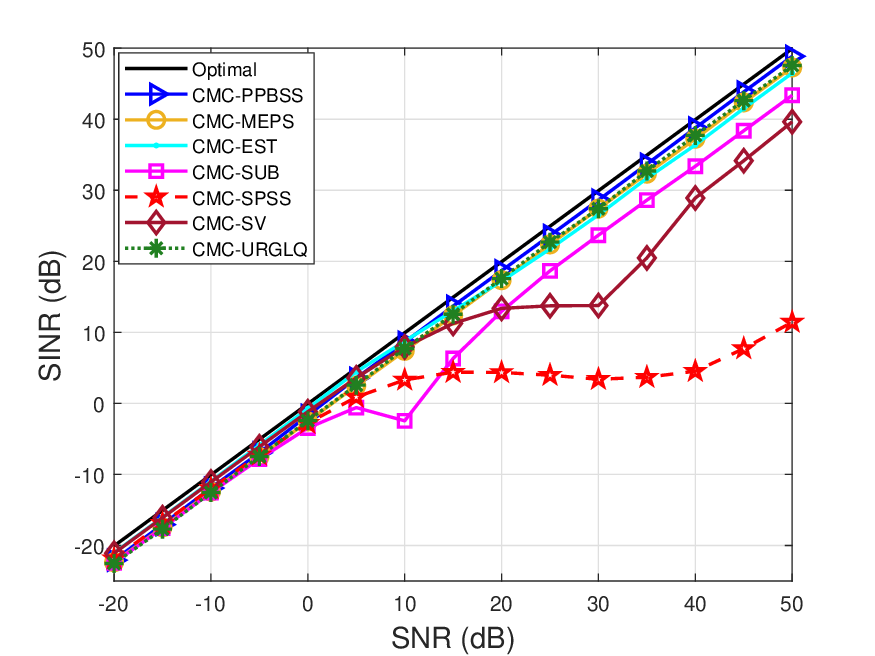}
\vspace{-0.07em}
	\caption{Output SINR versus input SNR in the case of look direction errors}
	\label{Look-SNR}
\end{figure}

The first example considers the impact of random signal look direction mismatch. We assume that the random direction mismatches of the desired signal and the interferers are uniformly distributed in $ [-4^\circ,4^\circ] $. This means that the actual SOI DoA is uniformly distributed in $ [-4^\circ, 4^\circ] $, and the DoAs of the interferers are uniformly distributed in $ [26^\circ,34^\circ] $ and $ [44^\circ, 54^\circ] $. Note that the DoAs of the desired signal and interferences change from run to run while remaining constant over samples. \\ \indent Fig.~\ref{Look-SNR} demonstrates that CMC-PPBSS, {CMC-URGLQ}, CMC-MEPS, and CMC-EST techniques closely follow the Optimal curve, indicating their robustness and efficiency. In contrast, while performing reasonably well, CMC-SUB shows a slight deviation; at 40 dB SNR, its SINR drops to around 35 dB. More notably, CMC-SPSS and CMC-SV exhibit significant performance degradation at higher SNR values. This indicates that these methods have scalability issues or robustness limitations in high SNR conditions, making them less suitable for high SNR scenarios. \\
\begin{figure}[!]
	\centering
	\includegraphics[height=2.5in]{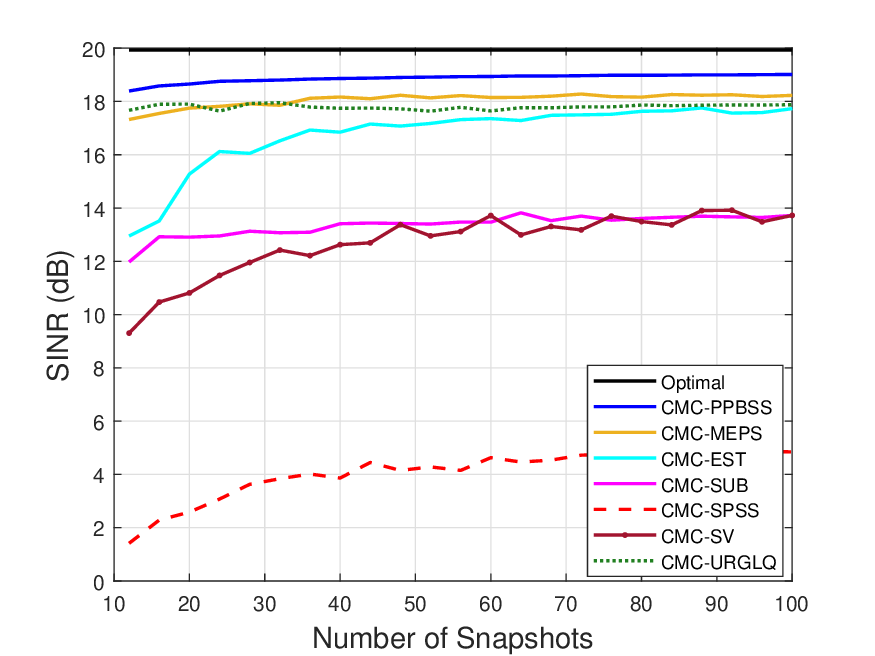}
 \vspace{-0.07em}
	\caption{Output SINR versus number of snapshots in the case of look direction errors} 
	\label{Look-Snapshots}
\end{figure}
\indent Fig.~\ref{Look-Snapshots} illustrates simulation results of SINR versus the number of snapshots for various methods. Among the methods, CMC-PPBSS, CMC-MEPS, and {CMC-URGLQ} demonstrate high and consistent performance, closely aligning with the Optimal curve, while the CMC-EST method's performance is increased by the increase in snapshots. In comparison, CMC-SUB performs moderately well, maintaining an SINR of around 12-14 dB. Notably, CMC-SPSS shows significantly lower SINR values, around 2-5 dB. These lower values indicate that CMC-SPSS and CMC-SV have limited effectiveness, but their performance is getting slightly better with the increased number of snapshots.

\subsection{Random signal look direction mismatch}
In this example, the impact of a random SV is considered. We assume that the SV of interferences and SOI are generated randomly as follows
\begin{align}
   \mathbf{a}(\theta) = \mathbf{a}(\hat{\theta}) + \mathbf{e}, \quad \theta \in (\theta_\mathrm{s}, \{ \theta_p \}_{p=1}^P)
\end{align}
where$\mathbf{a}(\hat{\theta})$ represents the assumed SV of the signal corresponding to $\hat{\theta}$, and $\mathbf{e}$ denotes the SV error, which is modeled as
\begin{align}
    \mathbf{e} = \frac{\epsilon}{\sqrt{L}} [e^{j \phi_1}, e^{j \phi_2}, \cdots, e^{j \phi_L}],
\end{align}
where $\epsilon = \| \mathbf{e} \|$ and satisfies the uniform distribution over $[0, \sqrt{0.3}]$; Also, $\{\phi_n \}_{n=1}^L$ are random phases with uniform distribution over  $[0, 2 \pi)$.
\begin{figure}[h]
	\centering
	\includegraphics[height=2.5in]{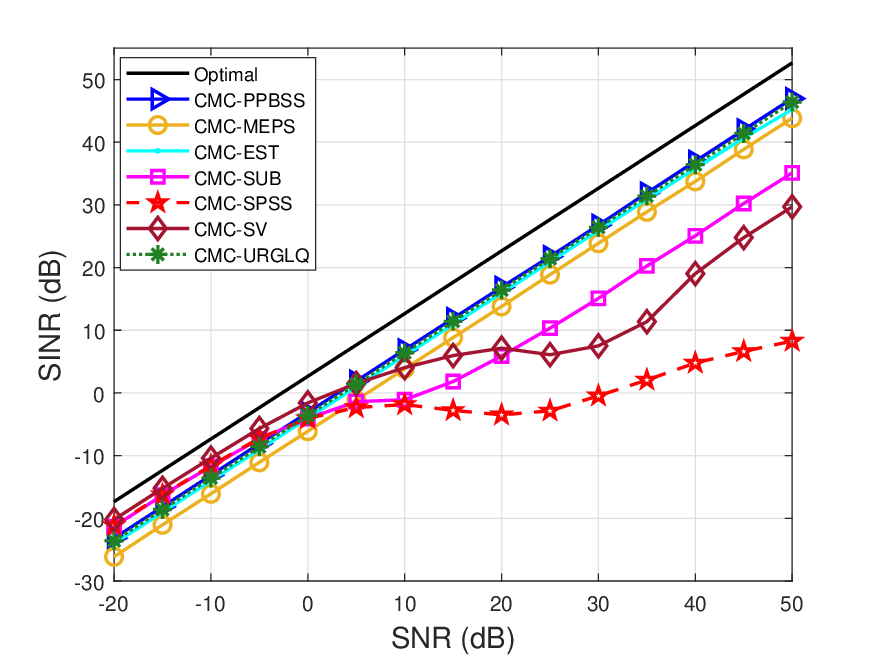}
\vspace{-0.05em}
	\caption{Output SINR versus input SNR in the case of SV random errors}
	\label{RV-SNR}
\end{figure}
\begin{figure}[h]
	\centering
	\includegraphics[height=2.5in]{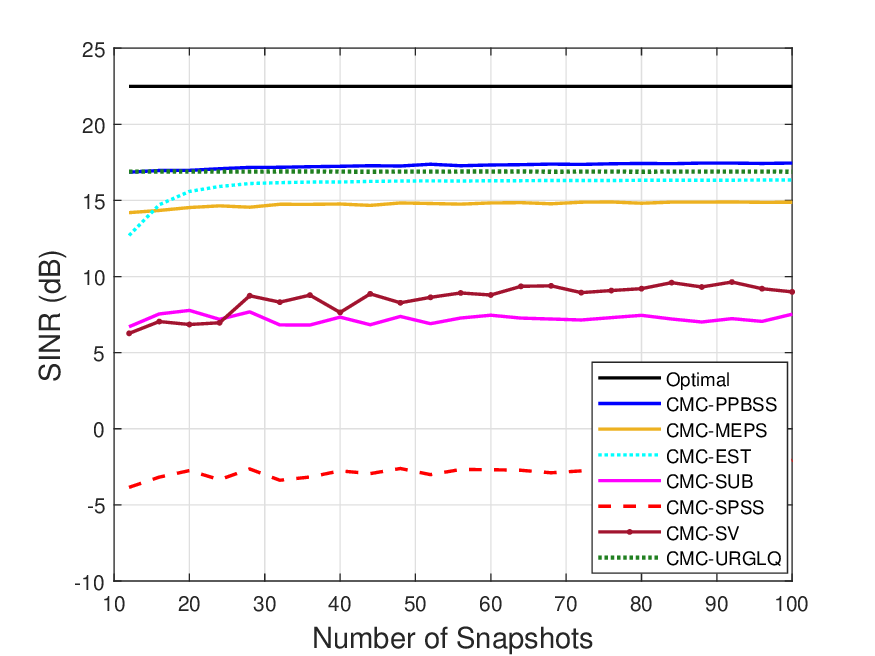}
\vspace{-0.05em}
	\caption{Output SINR versus number of snapshots in the case of SV random errors}
	\label{RV-Snapshots}
\end{figure}

Fig.~\ref{RV-SNR} demonstrates the output SINR of the beamformers versus the input SNR. It can be seen that due to the SOI component present in the sample covariance matrix, the performance of the CMC-SV decreases significantly with the increase of input SNR. Additionally, as the interference power is strong, the performance of the CMC-SUB is not perfect when random interference look direction errors are present. In contrast, the proposed CMC-PPBSS method can perform better than the reconstruction-based methods CMC-EST { and CMC-URGLQ}, while the CMC-MEPS performance is degraded compared to the proposed method. \\
\indent Fig.~\ref{RV-Snapshots} illustrates the performance of the tested beamformer in a scenario where the number of snapshots varies for a fixed SNR = 20 dB. All the beamformers demonstrate stable performance over the number of snapshots. We notice that the performance of the CMC-MEPS has declined compared to the result of this technique in the look direction mismatch. However, the CMC-SV method has demonstrated better results than the look direction mismatch, suggesting it is more resilient to random signal error. The performance of CMC-SV and CMC-SUB improves slightly as the number of snapshots increases. The proposed method outperforms all of the tested algorithms.

\subsection{Mismatch due to gain and phase perturbations errors}
\begin{figure}[!]
	\centering
	\includegraphics[height=2.5in]{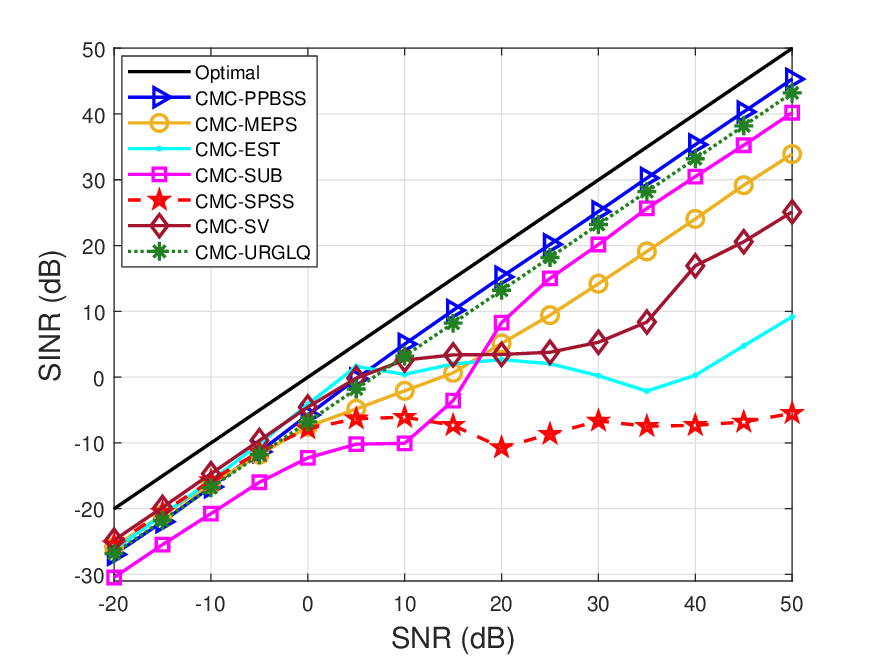}
\vspace{-0.05em}
	\caption{Output SINR versus input SNR in the case of phase and gain error}
	\label{PhaseGain-SNR}
\end{figure}

In this scenario, the impact of gain and phase perturbation errors on the array SV is examined under the assumption that the actual SV of the SOI includes look direction errors uniformly distributed within. $ \Delta \theta =  [-4^\circ 4^\circ]$. The calibration error is assumed to result from gain and phase perturbations in each sensor, which are represented as independent and identically distributed Gaussian random variables, given by $ \text{Ga} = \mathcal{N}(0,0.05^2)$ and $ \text{Pha} = \mathcal{N}(0,(0.025 \pi)^2)$ as
\begin{align}
    \mathbf{a}(\theta) = (1 + \text{Ga}) e^{j\text{Pha}}  e^{-j
2\pi (L-1)\bar{d}\sin(\theta + \Delta \theta)}.
\end{align}
\indent The output SINR versus the input SNR is depicted in Fig.~\ref{PhaseGain-SNR}. The CMC-PPBSS method demonstrates a high performance by closely following the optimal SINR. It is evident that the proposed method is robust against the gain and phase mismatches. The CMC-PPBSS method remains robust under gain-phase errors because it primarily relies on accurate angle information rather than precise gain and phase. By reconstructing the interference-plus-noise covariance matrix using angular sectors around the estimated DoAs in each snapshot, the method adapts itself to the change of DoA over time. Also, it avoids relying on potentially corrupted gain and phase data. This approach ensures that angular accuracy is preserved even in the presence of gain and phase errors, allowing for effective interference suppression and robust beamforming performance. Note that CMC-PPBSS {and CMC-URGLQ}  outperform other methods for SNR values greater than 10 dB. Conversely, CMC-SV demonstrates more resilience and performs better in low SNR conditions. However, the CMC-EST algorithm is very sensitive to compound errors and degrades performance with increasing SNR. Although CMC-SUB outperforms CMC-SPSS and CMC-SV, it still falls behind the top performers, maintaining an SINR of around 30 dB at an SNR of 10 dB. Notably, CMC-SPSS exhibits the poorest performance at high SNR values. In this example, when array perturbations are considered, most reconstruction-based methods experience a decrease in SINR at high SNR. This occurs because the inaccurate estimates of the interference SV in the reconstructed IPNC matrix limit the beamformers' ability to mitigate interference.\\
\begin{figure}[!]
	\centering
	\includegraphics[height=2.5in]{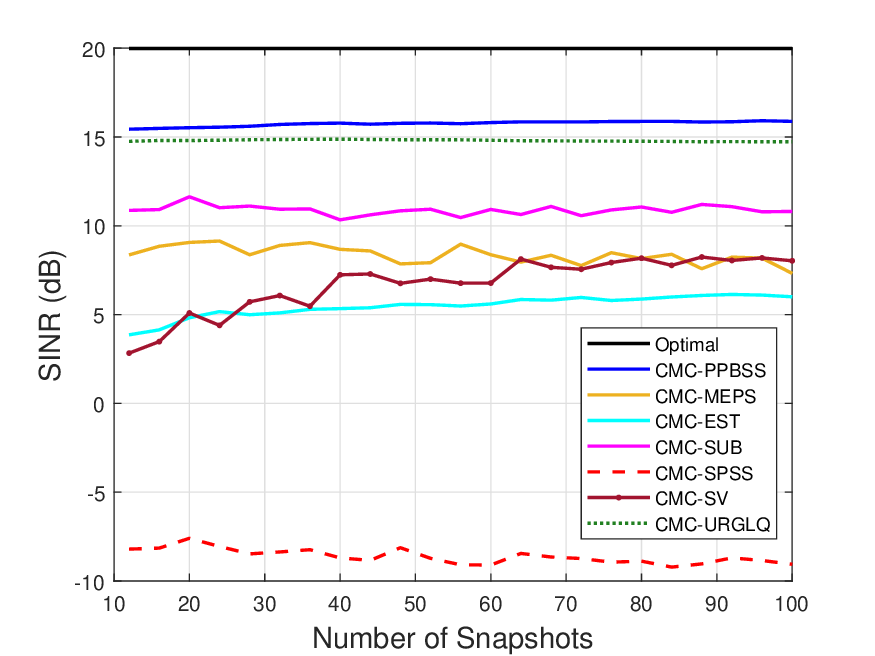}
\vspace{-0.05em}
	\caption{Output SINR versus number of snapshots in the case of phase and gain error}
	\label{PhaseGain-Snapshots}
\end{figure}
\indent The performance of the beamformers against the number of snapshots is illustrated in Fig.~\ref{PhaseGain-Snapshots}. In this uncalibrated array scenario, the performance of the CMC-EST reconstruction method degrades compared to the random signal and gain-phase mismatches due to imprecise prior knowledge of the SV. The algorithm in CMC-SUB utilizes the information about the uncertainty set regarding the SV. Moreover, the degradation of CMC-SPSS occurs because the angular sector corresponding to the DoA of interference in this method is not adjusted to account for variations in \(\Delta \theta\).

\begin{figure}[!]
	\centering
	\includegraphics[height=2.5in]{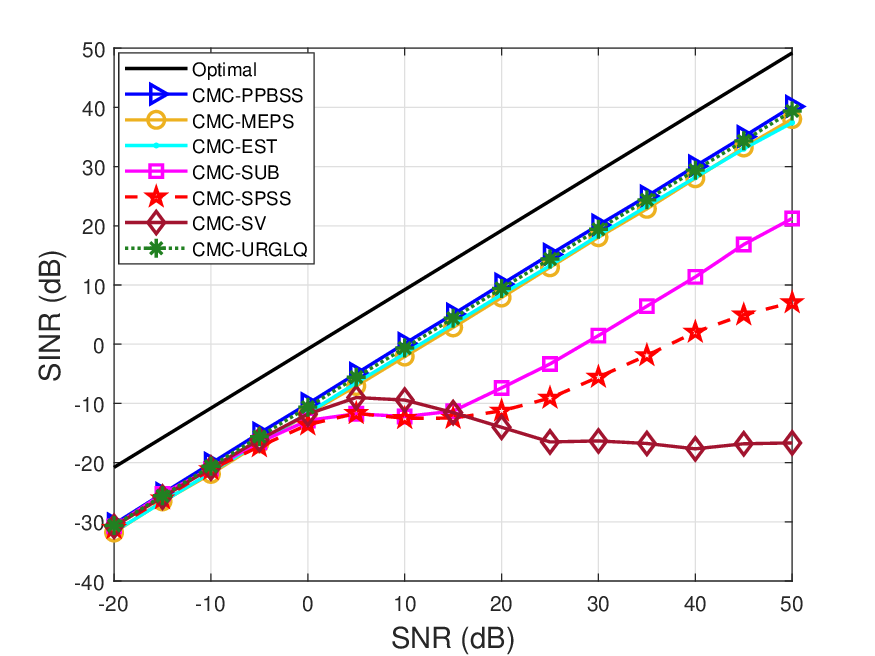}
\vspace{-0.05em}
	\caption{Output SINR versus input SNR in the case of array geometry mismatch}
	\label{Geometry-SNR}
\end{figure}

\begin{figure}[!]
	\centering
	\includegraphics[height=2.5in]{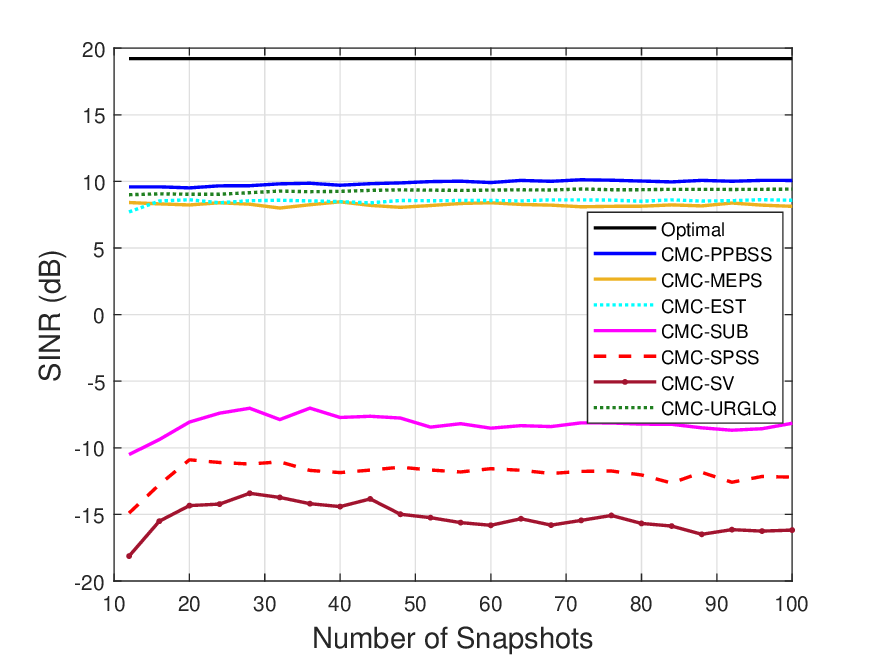}
\vspace{-0.05em}
	\caption{Output SINR versus number of snapshots in the case of array geometry mismatch}
	\label{Geometry-Snapshots}
\end{figure}

{\subsection{Array geometry mismatch }
In this example, we examine the impact of array geometry errors on the output SINR of various beamformers. Specifically, we consider perturbations in the sensor element positions, modeled as random errors uniformly distributed in the range $[-0.05, 0.05]$ of wavelengths. Fig.~\ref{Geometry-SNR} illustrates the output SINR versus input SNR for all beamformers, assuming a fixed number of snapshots $K = 50$. As can be seen, the proposed CMC-PPBSS method consistently outperforms the other beamformers across all SNR levels. The CMC-URGLQ and CMC-EST methods achieve performance slightly worse than the CMC-PPBSS method. In contrast, the CMC-SPSS method does not exhibit significant sensitivity to such mismatches, while the performance of the CMC-SV beamformer does not improve as the input SNR increases.\\
\indent Fig.~\ref{Geometry-Snapshots} illustrates the output SINR performance of various robust beamforming algorithms as a function of the number of snapshots under a fixed input SNR of 20 dB. The optimal beamformer performance serves as the upper performance bound. Among the tested methods, CMC-PPBSS, CMC-MEPS, and CMC-EST consistently achieve high SINR values, demonstrating robustness and effectiveness even with a limited number of snapshots. The CMC-URGLQ method also performs well, though with a slight performance gap compared to the top three methods. In contrast, the CMC-SUB algorithm shows moderate performance, gradually improving as the number of snapshots increases, but still significantly below that of the proposed method. Meanwhile, CMC-SPSS and CMC-SV exhibit the poorest performance, showing a high sensitivity to mismatch conditions. Overall, the results clearly indicate the superiority of the proposed or enhanced methods in maintaining reliable beamforming performance under practical conditions with snapshot and geometry uncertainties.}


\section{Conclusion}
This paper presented a preprocessing-based spatial sampling (PPBSS) method to design robust adaptive beamformers by addressing model errors and enhancing robustness. The key point was the PPBSS concept, which avoids estimating interference signals' power and SV. The IPNC matrix can be reconstructed using a generalized linear combination of the estimated and identity matrices, minimizing the MSE between the theoretical and estimated IPNC matrices. The angular sector of the desired signal is then used to construct the corresponding covariance matrix. The power and SV of the desired signal are estimated using the power method, which iteratively computes dominant eigenvalues and eigenvectors. This approach avoids estimating the noise covariance matrix, simplifying the process and improving robustness. The simulation results showed that the proposed method is more robust to various mismatches than state-of-the-art robust beamformers. 

\section*{Appendix A: Proof of \eqref{MSE}} \label{MSE proof}
\renewcommand\theequation{A.\arabic{equation}}
\setcounter{equation}{0}
To find \( \text{MSE} = \mathbb{E} \left\{ \|\eta \mathbf{I}-(1-\rho) \mathbf{R}_\mathrm{i+n}+\rho (\mathbf{C}-\mathbf{R}_\mathrm{i+n}) \|^2\right\}  \), let us break it into several well-defined steps:

1. Expand the Norm Squared:\\
Using the property of vector norms, the squared norm expands as:
\begin{align}\label{A+B}
\| \mathbf{A} + \mathbf{B} \|^2 = \| \mathbf{A} \|^2 + \| \mathbf{B} \|^2 + 2 \mathrm{Re}\{\langle \mathbf{A}, \mathbf{B} \rangle\},
\end{align}
where \(\langle \mathbf{A}, \mathbf{B} \rangle\) denotes the inner product. Here, let:
\begin{align}
\mathbf{A} =  \eta \mathbf{I}-(1-\rho) \mathbf{R}_\mathrm{i+n}, \quad \mathbf{B} = \rho_{} (\mathbf{C}-\mathbf{R}_\mathrm{i+n}).
\end{align}
We substitute \(\mathbf{A}\) and \(\mathbf{B}\) into \eqref{A+B}
\begin{align}
    \|\eta \mathbf{I}-&(1-\rho) \mathbf{R}_\mathrm{i+n}+\rho (\mathbf{C}-\mathbf{R}_\mathrm{i+n}) \|^2  = \nonumber \\ &\| \eta \mathbf{I}-(1-\rho) \mathbf{R}_\mathrm{i+n} \|^2 + \| \rho (\mathbf{C}-\mathbf{R}_\mathrm{i+n}) \|^2 \nonumber \\ &+ 2 \Re \left\{ \langle \eta \mathbf{I}-(1-\rho) \mathbf{R}_\mathrm{i+n} ,\rho (\mathbf{C}-\mathbf{R}_\mathrm{i+n}) \rangle \right\}.
\end{align}
2. Expectations of each term are given by
\begin{align}
    \mathbb{E} \{ \|\eta \mathbf{I}-&(1-\rho) \mathbf{R}_\mathrm{i+n}+\rho (\mathbf{C}-\mathbf{R}_\mathrm{i+n}) \|^2 \} =  \nonumber \\  &\| \eta \mathbf{I}-(1-\rho) \mathbf{R}_\mathrm{i+n} \|^2 + \mathbb{E} \{ \| \rho (\mathbf{C}-\mathbf{R}_\mathrm{i+n}) \|^2 \nonumber \\ &+ 2 \Re \left\{\langle \eta \mathbf{I}-(1-\rho) \mathbf{R}_\mathrm{i+n} ,\rho (\mathbf{C}-\mathbf{R}_\mathrm{i+n}) \rangle \right\} \}.
\end{align}
\rem{Since $\eta \mathbf{I}-(1-\rho) \mathbf{R}_\mathrm{i+n}$ and $\rho (\mathbf{C}-\mathbf{R}_\mathrm{i+n})$ are assumed independent, so}
{It can be numerically shown that inner product expectation is significantly smaller—by at least one or more orders of magnitude—than the MSE }
{\begin{align}
  \frac{\mathbb{E} \{  2 \Re \left( \langle \eta \mathbf{I}-(1-\rho) \mathbf{R}_\mathrm{i+n} ,\rho (\mathbf{C}-\mathbf{R}_\mathrm{i+n}) \rangle \right)  \}}{\text{MSE}}   \ll  1.
\end{align}}
and {the impact on the overall performance is negligible. Therefore, this term can be ignored in the analysis and optimization process without any noticeable loss of accuracy.} Using this, we have
\begin{align} \label{exp}
    \mathbb{E} \{ \|\eta \mathbf{I}-&(1-\rho) \mathbf{R}_\mathrm{i+n}+\rho (\mathbf{C}-\mathbf{R}_\mathrm{i+n}) \|^2 \} = \nonumber \\  &\| \eta \mathbf{I}-(1-\rho) \mathbf{R}_\mathrm{i+n} \|^2  + \mathbb{E} \{ \| \rho (\mathbf{C}-\mathbf{R}_\mathrm{i+n}) \|^2  \}.
\end{align}
For the first term, we have \( \| \eta \mathbf{I} - (1-\rho) \mathbf{R}_\mathrm{i+n} \|^2 \). Expanding the norm yields:
\begin{align} \label{norm}
\| \eta \mathbf{I}-&(1-\rho) \mathbf{R}_\mathrm{i+n} \|^2 \nonumber \\ & =  \text{Tr} \Big[ (\eta \mathbf{I}-(1-\rho) \mathbf{R}_\mathrm{i+n})^\mathrm{H} (\eta \mathbf{I}-(1-\rho) \mathbf{R}_\mathrm{i+n}) \Big] \nonumber \\
& = \text{Tr} \Big[ \eta^2 \mathbf{I} - 2\eta (1-\rho) \mathbf{R}_\mathrm{i+n}   + (1-\rho)^2 \mathbf{R}_\mathrm{i+n}^\mathrm{H} \mathbf{R}_\mathrm{i+n} \Big] \nonumber \\ 
& = \eta^2 L -2 \eta (1-\rho) \text{Tr} (\mathbf{R}_\mathrm{i+n}) + (1-\rho)^2\text{Tr} (\mathbf{R}_\mathrm{i+n}^\mathrm{H} \mathbf{R}_\mathrm{i+n}) \nonumber \\
& = \eta^2 L -2 \eta (1-\rho) \text{Tr} (\mathbf{R}_\mathrm{i+n}) + (1-\rho)^2 \| \mathbf{R}_\mathrm{i+n}\|^2,
\end{align}
where we have used the properties of trace as $ \text{Tr} (\mathbf{R}_\mathrm{i+n}^\mathrm{H} \mathbf{R}_\mathrm{i+n}) = \| \mathbf{R}_\mathrm{i+n}\|^2 $.
By replacing \eqref{norm} into \eqref{exp} we will have 
\begin{align}
    \text{MSE} = &\eta^2 L -2 \eta (1-\rho) \text{Tr} (\mathbf{R}_\mathrm{i+n}) \nonumber \\ &+ (1-\rho)^2 \| \mathbf{R}_\mathrm{i+n}\|^2 + \mathbb{E} \{ \| \rho (\mathbf{C}-\mathbf{R}_\mathrm{i+n}) \|^2\}.
\end{align}
This is the end of proof for \eqref{MSE}.

\section*{Appendix B: Proof of \eqref{zeta0}} \label{appendixB}
\renewcommand\theequation{B.\arabic{equation}}
\setcounter{equation}{0}
In order to find $\zeta$, we consider the following convex combination 
\begin{align} \label{second Problem}
    \tilde{\mathbf{R}}_\mathrm{i+n} = \eta \mathbf{I} + (1-\eta) \hat{\mathbf{R}}, \quad \eta \in (0,1).
\end{align}
The MSE of \eqref{second Problem} can be written as
\begin{align} \label{MSE for zeta}
\mathbb{E} \{ \| \tilde{\mathbf{R}}_\mathrm{i+n}-& \mathbf{R}_\mathrm{i+n} \|^2 \} =  \mathbb{E} \{ \| \eta(\mathbf{I}-\hat{\mathbf{R}})+(\hat{\mathbf{R}}-\mathbf{R}_\mathrm{i+n})\|^2 \} \nonumber \\
&=  \text { con. }+\eta^2 \mathbb{E} \{ \|\mathbf{I} -\hat{\mathbf{R}} \|^2 \} \nonumber \\
& -2 \eta \Re \{\operatorname{Tr}\left[\mathbb{E} \{(\hat{\mathbf{R}}-\mathbf{R}_\mathrm{i+n})(\hat{\mathbf{R}}-\mathbf{I})^* \} \right] \} \nonumber \\
= & \text { con. }+\eta^2 \mathbb{E} \{\|\mathbf{I}-\hat{\mathbf{R}}\|^2 \} -2 \eta \mathbb{E} \{ \| \hat{\mathbf{R}}-\mathbf{R}_\mathrm{i+n} \|^2 \},
\end{align}
where \text{con.} represents the constant number. Moreover, we have used the fact that $\hat{\mathbf{R}}$ is an unbiased estimate ($\mathbb{E} \{ \hat{\mathbf{R}} \} = \mathbf{R}_\mathrm{i+n}$). The unconstrained minimization of \eqref{MSE for zeta} for $\eta$ is given as
\begin{align} \label{etazero}
    \eta_0 = \frac{\mathbb{E} \{ \| \hat{\mathbf{R}}-\mathbf{R}_\mathrm{i+n} \|^2 \} }{\mathbb{E} \{ \| \hat{\mathbf{R}}-\mathbf{I} \|^2 \} } = \frac{\mathbb{E} \{ \| \hat{\mathbf{R}}-\mathbf{R}_\mathrm{i+n} \|^2 \}}{\mathbb{E} \{ \| \hat{\mathbf{R}}-\mathbf{R}_\mathrm{i+n} \|^2 \} + \| \mathbf{R}_\mathrm{i+n} -\mathbf{I} \|^2}.
\end{align}
It can be seen that $\eta_0$ is the solution of the minimization, and it falls into the interval (0,1). In the following, we show the algorithm in which $\eta_0$ could be computed using an estimate of $\mathbb{E} \{ \| \hat{\mathbf{R}}-\mathbf{R}_\mathrm{i+n} \|^2 \}$. Let define 
\begin{align}
    \hat{\mathbf{r}}_l = \frac{1}{K} \sum_{t=1}^K \mathbf{x}(t) x_l^*(t), \quad \mathbf{r}_l = \mathbb{E} \{ \mathbf{x}(t)x_l^*(t) \},
\end{align}
where $\hat{\mathbf{r}}_l$ and $\mathbf{r}_l$ are the $l$-th columns of $\hat{\mathbf{R}}$ and $\mathbf{R}_\mathrm{i+n}$, respectively. Also, $x_l(t)$ denotes the $l$-th element of $\mathbf{x}(t)$. Then, we can write
\begin{align}
      \mathbb{E} \{ \| \hat{\mathbf{R}}-\mathbf{R}_\mathrm{i+n} \|^2 \} = \sum_{l=1}^L \mathbb{E} \{ \| \hat{\mathbf{r}}_l - \mathbf{r}_l \|^2 \}.
\end{align}
By defining $\zeta \triangleq \mathbb{E} \{ \| \hat{\mathbf{R}}-\mathbf{R}_\mathrm{i+n} \|^2 \}$, now, we need to estimate $\mathbb{E} \{ \| \hat{\mathbf{r}}_l - \mathbf{r}_l \|^2 \} = \mathbb{E} \{ \| (1/K) \sum_{t=1}^K \mathbf{y}(t) - \mathbf{m} \|^2 \}$, where $\mathbf{y}(t) = \mathbf{x}(t) x_l^*(t)$ and $\mathbf{m}$ is the mean of $\mathbf{y}(t)$. The calculation can be given as
\begin{align} \label{calculation}
\mathbb{E} \{ \| \hat{\mathbf{r}}_l - \mathbf{r}_l \|^2 \} 
& =\frac{1}{K^2} \sum_{t=1}^K \sum_{\tilde{t}=1}^K \mathbb{E} \{[\mathbf{y}(t)-\mathbf{m}]^*[\mathbf{y}(t)-\mathbf{m}] \}  \nonumber \\
& =\frac{1}{K} \mathbb{E} \{\|\mathbf{y}(t)-\mathbf{m}\|^2\}.
\end{align}
The variance $\mathbb{E} \{\|\mathbf{y}(t)-\mathbf{m}\|^2\}$ in \eqref{calculation} can be estimated as
\begin{align}
    \frac{1}{K} \sum_{t=1}^K\|\mathbf{y}(t)-\hat{\mathbf{m}}\|^2 ; \quad \hat{\mathbf{m}}=\frac{1}{K} \sum_{t=1}^K \mathbf{y}(t)=\hat{\mathbf{r}}.
\end{align}
It follows that
\begin{align}
    \mathbb{E} \{ \| \hat{\mathbf{r}}_l - \mathbf{r}_l \|^2 \} &= \frac{1}{K^2} \sum_{t=1}^K \| \mathbf{x}(t)x_l^*(t) - \hat{\mathbf{m}} \|^2 \nonumber \\
    & = \frac{1}{K^2} \sum_{t=1}^K \sum_{l=1}^L \| \mathbf{x}(t)x_l^*(t) - \hat{\mathbf{m}} \|^2 \nonumber \\
    & = \frac{1}{K} \sum_{l=1}^L \frac{1}{K} \sum_{t=1}^K \| \mathbf{x}(t)x_l^*(t) - \hat{\mathbf{m}} \|^2 \nonumber \\
    & = \frac{1}{K} \sum_{l=1}^L \left[ \frac{1}{K} \sum_{t=1}^K \| \mathbf{x}(t) \|^2  |x_l(t) |^2 - \| \hat{\mathbf{r}}_l \|^2 \right]  \nonumber \\
    & = \frac{1}{K^2} \sum_{t=1}^K \| \mathbf{x}(t)\|^4 -  \frac{1}{K} \|\hat{\mathbf{R}} \|^2  = \hat{\zeta}.
\end{align}
This is the end of the proof for \eqref{zeta0}.
\section*{Appendix C: definition of  Pearson correlation} \label{appendixC}
\renewcommand\theequation{C.\arabic{equation}}
\setcounter{equation}{0}
For two matrices $ \mathbf{X} \in \mathbb{R}^{n \times p} $ and $ \mathbf{Y} \in \mathbb{R}^{n \times q} $, we can calculate the correlation coefficient for each pair of variables $ x $ and $ y $ as follows
\begin{align}
    \text{corr}(x, y) = \frac{\text{Cov}(x, y)}{\sigma_x \sigma_y}, \tag{C.1}
\end{align}
where  $ \text{Cov}(x, y) = \frac{1}{n-1} \sum_{i=1}^n (x_i - \bar{x})(y_i - \bar{y}) $ is the covariance and
 $ \sigma_x = \sqrt{\frac{1}{n-1} \sum_{i=1}^n (x_i - \bar{x})^2} $ is the standard deviation of $ x $ while  $ \bar{x} $ and $ \bar{y} $ are the means of $ x $ and $ y $, respectively.
For matrix $ \mathbf{X} $, the output is a $ p \times p $ correlation matrix:
\begin{align} \label{Rij}
R(i,j) = \text{corr}(\mathbf{X}(:,i), \mathbf{X}(:,j)), \tag{C.2}
\end{align}
where $ \mathbf{X}(:,i) $ denotes the $ i $-th column of $ \mathbf{X} $. If two matrices $ \mathbf{X} $ and $ \mathbf{Y} $ are provided, \eqref{Rij} computes the cross-correlation between columns of $ \mathbf{X} $ and $ \mathbf{Y} $.

\bibliographystyle{IEEEtran}
\bibliography{Reference}
\end{document}